\documentclass[%
twocolumn, 
pre,
superscriptaddress,
 aps,
floatfix,
]{revtex4-2}
\usepackage{graphicx}
\usepackage{amssymb}
\usepackage{color}
\usepackage{amsmath}
\usepackage[table,xcdraw]{xcolor}
\usepackage{mathrsfs}
\usepackage[bookmarks=true,colorlinks,urlcolor=blue,linkcolor=blue,anchorcolor=blue,citecolor=blue,unicode]{hyperref}
\usepackage{bookmark}
\usepackage{bm}
\usepackage{chemformula}
\usepackage{footnote}
\graphicspath{{Figures/}}

\begin{document}
\preprint{Phys. Rev. E}

\title{Quantifying the temporal stability of international fertilizer trade networks}
\author{Mu-Yao Li}
  \affiliation{School of Business, East China University of Science and Technology, Shanghai 200237, China} %
  \affiliation{Research Center for Econophysics, East China University of Science and Technology, Shanghai 200237, China} %

\author{Li Wang}%
  \affiliation{Research Center for Econophysics, East China University of Science and Technology, Shanghai 200237, China} %
  \affiliation{School of Physics, East China University of Science and Technology, Shanghai 200237, China} %

\author{Wen-Jie Xie}%
  \email{wjxie@ecust.edu.cn}
  \affiliation{School of Business, East China University of Science and Technology, Shanghai 200237, China} %
  \affiliation{Research Center for Econophysics, East China University of Science and Technology, Shanghai 200237, China} %

\author{Wei-Xing Zhou}
  \email{wxzhou@ecust.edu.cn}
  \affiliation{School of Business, East China University of Science and Technology, Shanghai 200237, China} %
  \affiliation{Research Center for Econophysics, East China University of Science and Technology, Shanghai 200237, China} %
 \affiliation{School of Mathematics, East China University of Science and Technology, Shanghai 200237, China} %

\date{\today}
\begin{abstract}
The importance of fertilizers to agricultural production is undeniable, and most economies rely on international trade for fertilizer use. The stability of fertilizer trade networks is fundamental to food security. We use three valid methods to measure the temporal stability of the overall network and different functional sub-networks of the three fertilizer nutrients N, P and K from 1990 to 2018. The international N, P and K trade systems all have a trend of increasing stability with the process of globalization. The large-weight sub-network has relatively high stability, but is more likely to be impacted by extreme events. The small-weight sub-network is less stable, but has a strong self-healing ability and is less affected by shocks. Overall, all the three fertilizer trade networks exhibit a stable core with restorable periphery. The overall network stability of the three fertilizers is close, but the K trade has a significantly higher stability in the core part, and the N trade is the most stable in the non-core part.
\end{abstract}

\maketitle


\section{Introduction}

The use of fertilizers is an important bridge for human activities to intervene in ecological activities \citep{Matson-Partonr-Power-Swift-1997-GenSyst}, and the stability of the human-dominated ecological environment \citep{Vitousek-1997-Science} is inevitably becoming more and more dependent on the stable supply of fertilizers from international trade. The instability of a certain fertilizer nutrient is passed directly into the cycle of the ecosystem \citep{Galloway-Townsend-Erisman-Bekunda-Cai-Freney-Martinelli-Seitzinger-Sutton-2008-Science}. At the same time, the supply of fertilizers based on international trade has been the underlying basis for food production in most economies. If there is a lack of fertilizers, a large number of economies will have to face the problem of food shortages due to grain production reductions of up to 50\% \citep{Stewart-Dibb-Johnston-Smyth-2005-AgronJ}. Even worse is that most of these economies do not have the financial resources to buy enough food directly on the international market to meet the food needs of their people \citep{Foley-Ramankutty-Brauman-Cassidy-Gerber-Johnston-Mueller-O'Connell-Ray-West-Balzer-Bennett-Carpenter-Hill-Monfreda-Polasky-Rockstrom-Sheehan-Siebert-Tilman-Zaks-2011-Nature}. Therefore, the stability of the fertilizer trade system is an important research object for ecosystem stability and food security.

Networks are widely used as a convenient quantitative representation of patterns of interactions between the constituents of complex systems. International trade, a natural network with economies as nodes and inter-economy trades as links, has attracted the attention of a large number of physicists and economists \citep{Hidalgo-Klinger-Barabasi-Hausmann-2007-Science,Hidalgo-Hausmann-2009-ProcNatlAcadSciUSA,Chaney-2014-AmEconRev}. In the past decade, researchers are shifting their attention from the static aggregate international-trade network (ITN) to more complex network models to study more practical problems. Two main representation approaches have been employed to address this issue. One is multi-layer networks \citep{Barigozzi-Fagiolo-Garlaschelli-2010-PhysRevE,Mastrandrea-Squartini-Fagiolo-Garlaschelli-2014-PhysRevE,Cimini-Squartini-Saracco-Garlaschelli-Gabrielli-Caldarelli-2019-NatRevPhys}. International trade contains more than 6,000 classified commodities. In a multi-layer network structure, the trade of each commodity can form a layer of a multi-layer network, and each commodity is related to each other \citep{Hidalgo-Klinger-Barabasi-Hausmann-2007-Science}. Another is temporal networks. Temporal networks can be considered as a special kind of multi-layer network \citep{Boccaletti-Bianconi-Criado-delGenio-GomezGardenes-Romance-SendinaNadal-Wang-Zanin-2014-PhysRep}, data at different times constitutes different layers in the network. Benefiting from the long-term stability of the economies, the international trade network with available data spanning up to 30 years is naturally suitable to the temporal network model. The temporal network setup \citep{Holme-Saramaki-2012-PhysRep} is characterized by analyzing the similarities and differences of 
the same object at different times, which allows us to explore the stability of the system from the dimension of time \citep{Petri-Expert-2014-PhysRevE,Sun-Gao-Zhong-Liu-2017-PhysicaA}, named temporal stability.
%

The stability of the static network structure can be defined as the structural robustness of a complex system to perturbation \citep{Albert-Jeong-Barabasi-2000-Nature}. It has extensive research and applications in ecological networks \citep{Montoya-Pimm-Sole-2006-Nature}, protein networks \citep{Maslov-Sneppen-2002-Science}, internet networks \citep{Cohen-Erez-benAvraham-Havlin-2000-PhysRevLett}, interbank networks \citep{Boss-Elsinger-Summer-Thurner-2004-QuantFinanc}, food-web networks \citep{Dunne-Williams-Martinez-2002-EcolLett}, and so on. The stability of the static network defines certain simulation index of the static network structure. Researchers simulate the occurrence and impact of perturbation and differentiate networks by the degree of impact. Relatively, the temporal stability refers to the robustness of the complex system to the flow of time. A system with higher temporal stability has stronger time-invariant properties. The stability of communities in temporal networks is an important subject of considerable attention. Most of the research in this area is based on random walks, and there have been many excellent results \citep{Palla-Barabasi-Vicsek-2007-Nature, Mucha-Richardson-Macon-Porter-Onnela-2010-Science, Delvenne-Yaliraki-Barahona-2010-ProcNatlAcadSciUSA}. The stability of both the whole and part of the system is valuable for research \citep{Hand-2007-BCSDI}. Through quantitative research on the temporal stability of fertilizer trade, we aim to understand the overall resistance of the fertilizer trade system to time-varying changes. 

Estimating the temporal stability is essentially a comparison of two temporally consecutive networks of known node correspondence \citep{Tantardini-Ieva-Tajoli-Piccardi-2019-SciRep}. For this, scholars have summarized many methods, such as DeltaCon \citep{Koutra-Shah-Vogelstein-Gallagher-Faloutsos-2016-ACMTransKnowlDiscovData}, Cut distance, Euclidean distance, Manhattan distance, Canberra distance, and Jaccard index \citep{Hand-Mannila-Smyth-2001}. 
In order to illustrate the temporal stability from the physical meaning well, we investigate the stability of the temporal network from three perspectives. The first is the comprehensive probability that the network structure existing at moment $t$ inherits to moment $t+1$. Repeatable structure shows strong resistance to time. Here we introduce the structural inheritance index. 
The second is the correlation between the network structures at two moments, that is, the uncertainty of the network structure at $t+1$ that can be reduced by the network structure at $t$. The mutual information from information theory \cite{Schieber-Carpi-Frery-Rosso-Pardalos-Ravetti-2016-PhysLettA} will be used to characterize this part. The third is the similarity between structures of two networks, which we use the Jaccard index to represent \citep{Palla-Barabasi-Vicsek-2007-Nature}. So in this study, we will use the three indicators, structural inheritance, mutual information and Jaccard index, to quantify the temporal stability of international fertilizer trade networks from the three dimensions of inheritance, correlation and similarity. 

For fertilizers, regardless of their names or manufacturers, the key is always the content of the three main nutrients, nitrogen (expressed as \ch{N}), phosphorus (expressed as \ch{P2O5} and abbreviated as \ch{P}), and potash (expressed as \ch{K2O} and abbreviated as \ch{K}). These play the most important roles in agricultural production \citep{Huang-Gao-Lin-Cui-Zhong-Huang-2020-ResourConservRecycl,Wang-Miao-You-Ren-2021-ResourConservRecycl}. The international trade of the three fertilizer nutrients is an important part of the \ch{N} cycle, \ch{P} cycle, and \ch{K} cycle, which is related to ecological development  \citep{Reinhard-Planavsky-Gill-Ozaki-Robbins-Lyons-Fischer-Wang-Cole-Konhauser-2017-Nature,Lwin-Murakami-Hashimoto-2017-ResourConservRecycl} and food security \citep{Galloway-Townsend-Erisman-Bekunda-Cai-Freney-Martinelli-Seitzinger-Sutton-2008-Science,Liang-Yu-Kharrazi-Fath-Feng-Daigger-Chen-Ma-Zhu-Mi-Yang-2020-NatFood,Harder-Giampietro-Smukler-2021-ResourConservRecycl}, which is an issue of widespread concern to researchers and policy makers.

The rest of the paper is organized as follows. Section~\ref{S:Methods and data} describes the database and explains the methodology employed to build the international fertilizer trade networks. Section~\ref{S1:StructuralPatterns} reports the calculation and results of the three indicators. Conclusions are presented in Section~\ref{S:Conclusion}.

\section{Methods and data}
\label{S:Methods and data}
\subsection{Data}

The fertilizer trade data sets from 1990 to 2018 we used were retrieved from the UN Comtrade Database (https://comtrade.un.org). In the UN Comtrade Database, the six-digit HS codes represent the types of commodities. Compared with different fertilizer commodities, the three main nutrients (N, P and K) are the key to ecological and food research. So we calculate the nutrient content of each fertilizer commodity to obtain the trade networks of the N, P and K nutrients using the method proposed by the Food and Agriculture Organization (FAO, http://faostat.fao.org). The basic unit of fertilizer trade data is tonne. Since every economy reports their import and export data, we have preprocessed the repeated trade data \citep{Feenstra-Lipsey-Deng-Ma-Mo-2005-NBER}.

\subsection{Network construction}

If the amount of fertilizer nutrient $f$ exported from economy $i$ to economy $j$ in year $t$ is $w_{ij}^{f}(t)$ tons, then the temporal network of fertilizer trade can be expressed as
\begin{equation}
    {\mathbf{W}}^{f}(t) = \left[w_{ij}^{f}(t)\right],
    \label{Eq:Fertilizer:Wft}
\end{equation}
while $f\in\{\mathrm{N}, \mathrm{P}, \mathrm{K}\}$ is one of the three main nutrients of fertilizers.

Let $\mathscr{V}=\{1, 2, \cdots, n \}$ represents the set of different economies ({\it{i.e.}} nodes). The set of directed trade relationships ({\it{i.e.}} links) between economies is expressed as $\mathscr{E}=\{e_{ij}\}$, where $e_{ij}=1$ means economy $i$ exports to economy $j$ with the trade volume $w_{ij}$. If there is no trade relationship between the two economies, then $e_{ij}=e_{ji}=w_{ij}=w_{ji}=0$. There are no self-loops in the international fertilizer trade networks such that $e_{ii} = w_{ii} =0$ for all $i\in\mathscr{V}$. Overall, there are $n^2-n$ potential links.

\begin{figure*}[!ht]
    \centering
    \includegraphics[width=0.95\linewidth]{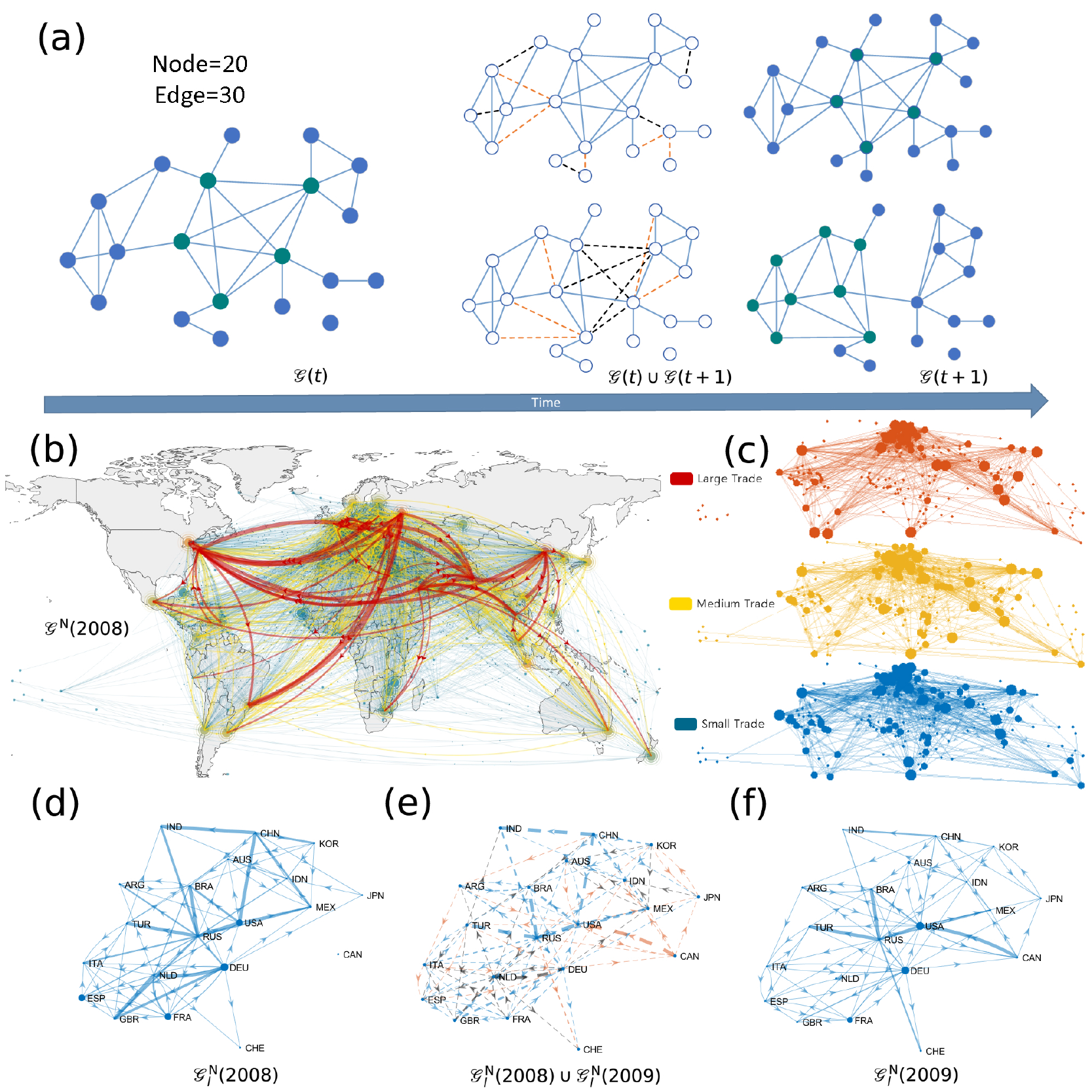}
    \caption{Structure and schematic dynamics of the N trade network. (a) Schematic diagram of a simple network evolving over time. A same-scale change in the adjacency matrix can lead to huge different structural change. Nodes marked in green are considered as the core group of the network. In $\mathscr{G}(t)\cup \mathscr{G}(t+1)$, the red dotted line represents the links that will be connected, and the black dotted line represents the links that will be broken. (b) The N trade network $\mathscr{G}^{\mathrm{N}}(2008)$ in 2008. The red lines represent the large trades while their widths corresponds to the trade volumes. The yellow lines represent the medium trades and the blue lines represent the small trades. (c) The three subgraphs of the N network, $\mathscr{G}_{l}^{\mathrm{N}}(2008)$, $\mathscr{G}_{m}^{\mathrm{N}}(2008)$ and $\mathscr{G}_{s}^{\mathrm{N}}(2008)$, corresponding to the three parts of high, medium and low weight respectively. The node size symbolizes the node's betweenness centrality. (d,e,f) The evolution process from $\mathscr{G}_{l}^{\mathrm{N}}(2008)$ to $\mathscr{G}_{l}^{\mathrm{N}}(2009)$. Only 20 large economies are drawn. The line width corresponds to the trade volume and the node size corresponds to economy's betweenness centrality.}
    \label{Fig:FertilizerNetrnetwork}
\end{figure*}

In Fig.~\ref{Fig:FertilizerNetrnetwork}(a), we characterize the evolution of a simple network from time $t$ to $t+1$. For the network, the impact of the same-scale link breakdown and occurrence in different parts is almost incomparable. The changes happened in the core part have far greater influence than that happened in the periphery structure. This illustrates the importance of quantifying temporal stability in multiple dimensions.

We display a map of the N trade network in 2008 $\mathscr{G}^{\mathrm{N}}(2008)$ in Fig.~\ref{Fig:FertilizerNetrnetwork}(b). The large trades are colored in red, while the widths correspond to the trade volumes. The medium trades are colored in yellow and the small trades in blue. Fig.~\ref{Fig:FertilizerNetrnetwork}(c) separates the links by their weights and presents them as three subgraphs. Large trades are defined as the top 20\% largest trades. Different from the unweighted network, in the actual N, P, K trade networks, the large-trade sub-network that occupies at least more than 90\% of the total trade volume is the most important part of the network. In order to ensure the same number of links in different subgraphs for comparability, we extract the links with the weights between 40\% and 60\% to form the medium-trade sub-network and the links with the weights less than 20\% to form the small-trade sub-network. We will study the stability of the three sub-networks in time-varying evolution for digging the internal structure. 

In Fig.~\ref{Fig:FertilizerNetrnetwork}(d,e,f), the evolution of the large-trade sub-network $\mathscr{G}_{l}^{\mathrm{N}}$ from 2008 to 2009 is depicted. The line width stands for the trade volume and the node size represents the economy's betweenness centrality, which is the number of shortest paths through that node. The line width in $\mathscr{G}_{l}^{\mathrm{N}}(2008)\cup \mathscr{G}_{l}^{\mathrm{N}}(2009)$ is calculated by the total weight of $\mathscr{G}_{l}^{\mathrm{N}}(2008)$ and $\mathscr{G}_{l}^{\mathrm{N}}(2009)$. To have clear pictures, we only plot the trade relations between the 20 largest economies according to the GDP in 2008. The calculation of line width and node size is based on the overall sub-network. 
From the evolution process, we can find that there is a relatively large change in the core structure of the network, due to the economic crisis. Most of the broken links are concentrated in the European and East Asian economies. Canada emerges as a big player with many new connections, who disconnected in 2008 because of the first wave of the economic crisis. These make the United States and Germany play a more important role in transportation and reproduction, while Russian and China still play the roles of main producer in their regions. Based on these information, 
we believe that diversely quantifying the temporal stability of trade networks can give us a deeper understanding of the evolution of trade systems.

\section{Stability of temporal networks}
\label{S1:StructuralPatterns}

In this section, we characterize the temporal stability of the N, P and K trade network and its three sub-network through three different methods. High stability means that the network is less prone to large-scale changes. The three methods, structural inheritance, mutual information and Jaccard index, will be used to quantify the temporal stability of the international fertilizer trade networks from the dimensions of inheritance, correlation and similarity. The large trades, medium trades and small trades will be separated to analyze the discrepancies in the stability of different functional parts of the networks. By comparing the results obtained from the three methods, an analysis of the temporal stability of the structure of the fertilizer trading system will be presented.

\subsection{Structural inheritance}

We consider two successive networks $\mathscr{G}(t)$ and $\mathscr{G}(t+1)$ with the same set ${\mathscr{V}}$ of vertices, but in general with different sets ${\mathscr{E}(t)}$ and ${\mathscr{E}(t+1)}$ of links. Let us indicate the number of vertices in both networks with $N_{\mathscr{V}}= \sharp\left[\mathscr{V}\right]$, the number of links in $\mathscr{G}(t)$ with $N_{\mathscr{E}(t)}= \sharp\left[\mathscr{E}(t)\right]$, and the number of links in $\mathscr{G}(t+1)$ with $N_{\mathscr{E}(t+1)}= \sharp\left[\mathscr{E}(t+1)\right]$, where $\sharp\left[\mathbf{X}\right]$ represents the cardinal number of set $\mathbf{X}$. The number $N_{{\mathscr{E}}(t)\cap {\mathscr{E}}(t+1)}$ of same links which are present in both networks is determined as follows
\begin{equation}
    N_{{\mathscr{E}}(t)\cap {\mathscr{E}}(t+1)} 
    = \sharp\left[{\mathscr{E}}_{t}\cap {\mathscr{E}}_{t+1}\right]
    = \sum_{i\in\mathscr{V}}\sum_{j\in\mathscr{V}}e_{ij}(t)e_{ij}(t+1),
\end{equation}
where $e_{ii}=0$ by definition. 

In order to evaluate the transfer situation between $\mathscr{G}(t)$ and $\mathscr{G}(t+1)$, we associate a binary variable $x_t$ with all ordered pair of vertices in the first network $\mathscr{G}(t)$ and a binary variable $x_{t+1}$ with all ordered pair of vertices in the second network $\mathscr{G}(t+1)$. Note that an ordered pair $(i,j)$ is different from $(j,i)$. The variable $x_t$ takes the value 1 if two vertices are linked in the network $\mathscr{G}(t)$ and it is 0 otherwise, while the variable $x_{t+1}$ takes the value 1 if two vertices are linked in the network $\mathscr{G}(t+1)$ and it is 0 otherwise. On this basis, we calculate the joint probabilities $p(x_t, x_{t+1})$ of the two variables $x_t\in\{0,1\}$ and $x_{t+1}\in\{0,1\}$. The joint probability $p(1,1)$ that an link $e_{ij}$ exists in both networks $\mathscr{G}(t)$ and $\mathscr{G}(t+1)$ is given by
\begin{equation}
  p(1,1) =\frac{N_{{\mathscr{E}}(t)\cap {\mathscr{E}}(t+1)}}{N(N-1)}.
\label{Eq:p11}
\end{equation}
The joint probability $p(1, 0)$ that an link $e_{ij}$ exists in $\mathscr{G}(t)$ but not in $\mathscr{G}(t+1)$ is given by
\begin{equation}
  p(1, 0) = \frac{N_{\mathscr{E}(t)}-N_{{\mathscr{E}}(t)\cap {\mathscr{E}}(t+1)}}{N(N-1)},
\label{Eq:p10}
\end{equation}
which is the probability of link disappearance in the transfer process. The joint probability $p(0, 1)$ that a link $e_{ij}$ does not exist in $\mathscr{G}(t)$ but exists in $\mathscr{G}(t+1)$ is given by
\begin{equation}
  p(0, 1) = \frac{N_{\mathscr{E}(t+1)}-N_{{\mathscr{E}}(t)\cap {\mathscr{E}}(t+1)}}{N(N-1)},
\label{Eq:p01}
\end{equation}
which is the probability of link appearance in the transfer process. The joint probability $p(0, 0)$ that a link $e_{ij}$ does not exist in neither networks $\mathscr{G}(t)$ and $\mathscr{G}(t+1)$ is given by
\begin{equation}
  p(0, 0) = 1-\frac{N_{\mathscr{E}(t)}+N_{\mathscr{E}(t+1)}-N_{{\mathscr{E}}(t)\cap {\mathscr{E}}(t+1)}}{N(N-1)}.
\label{Eq:p00}
\end{equation}
It is worth noting that
\begin{equation}
  p(1, 1)+p(1, 0)+p(0, 1)+p(0, 0)=1.
\end{equation}

Although we calculated the probabilities of the four kinds of link changes between the two networks, this does not objectively reflect the actual tendency of the links. For example, in two very dense networks, $p(1,1)$ will obviously have a higher value even if they are randomly distributed. In order to estimate the effect of random distribution, we calculate the probability of the existence of directed links between two random nodes in the network $\mathscr{G}(t)$,
\begin{equation}
  p_{\mathscr{G}(t)}(1) = \frac{N_{\mathscr{E}(t)}}{N(N-1)}
\label{Eq:p(1)}
\end{equation}
and the corresponding probability that there is no directed link between two random nodes
\begin{equation}
  p_{\mathscr{G}(t)}(0) = 1-p_{\mathscr{G}(t)}(1) = 1 - \frac{N_{\mathscr{E}(t)}}{N(N-1)}.
\label{Eq:p(0)}
\end{equation}
Then we assume that $\mathscr{G}^{'}(t)$ and $\mathscr{G}^{'}(t+1)$ are two independent random networks. The network density of $\mathscr{G}^{'}(t)$ is the same as $\mathscr{G}^(t)$, and the network density of $\mathscr{G}^{'}(t+1)$ is the same as $\mathscr{G}(t+1)$. Therefore, the probability $p_{\mathscr{G}^{'}(t)} (1)$ is the probability that a randomly selected ordered pair of vertices is linked in the network $\mathscr{G}^{'}(t)$
\begin{equation}
  p_{\mathscr{G}^{'}(t)}(1) = p_{\mathscr{G}(t)}(1) = \frac{N_{\mathscr{E}(t)}}{N(N-1)},
\end{equation}
which is the density of the directed network $\mathscr{G}(t)$. And the probability $p_{\mathscr{G}^{'}(t)}(0)$ that a randomly selected ordered pair of vertices is not linked in the network $\mathscr{G}^{'}(t)$ is
\begin{equation}
  p_{\mathscr{G}^{'}(t)}(0) = p_{\mathscr{G}(t)}(0) = 1 - \frac{N_{\mathscr{E}(t)}}{N(N-1)}.
\end{equation}

\begin{figure*}[!ht]
    \centering
    \includegraphics[width=0.49\linewidth]{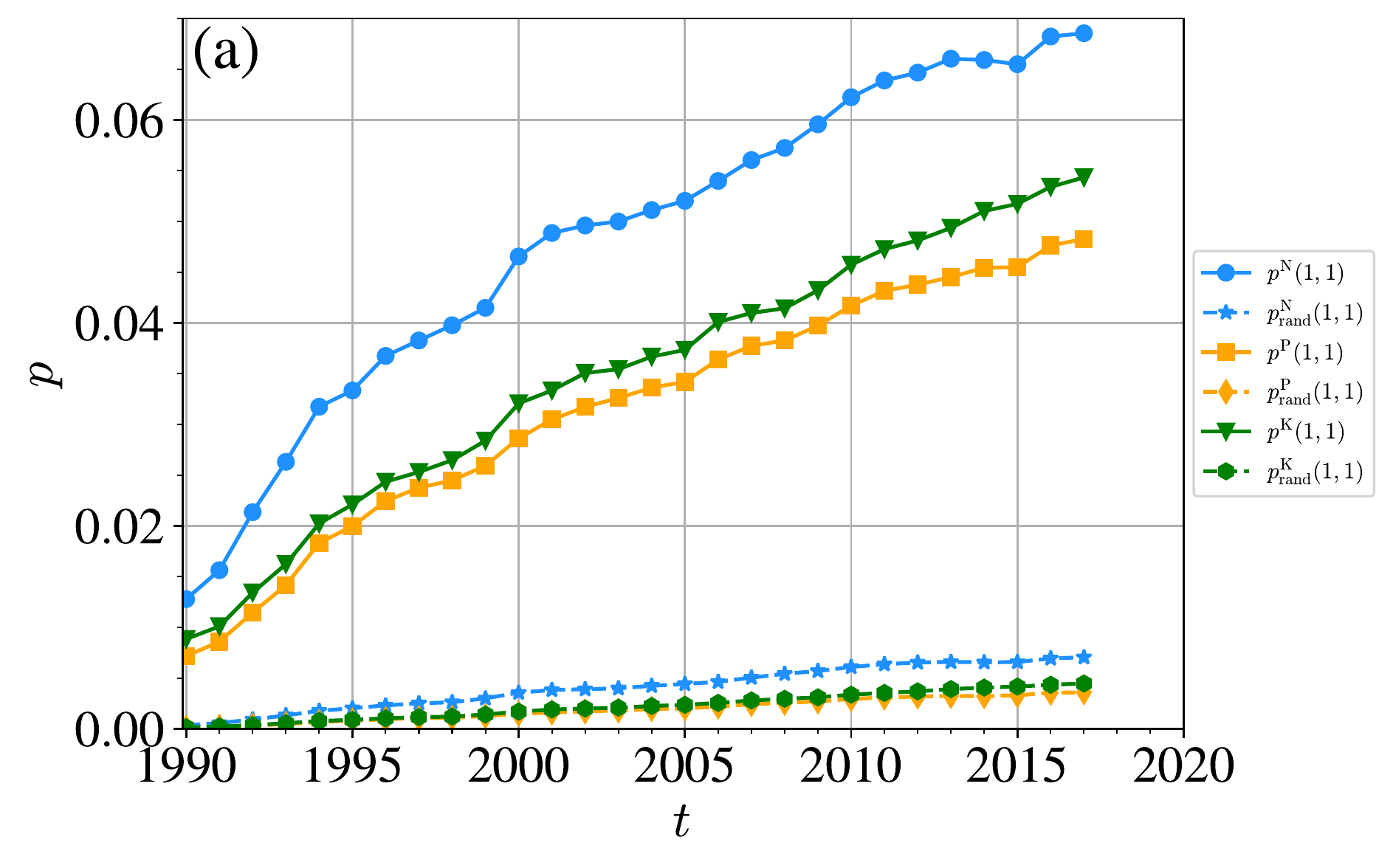}
    \includegraphics[width=0.49\linewidth]{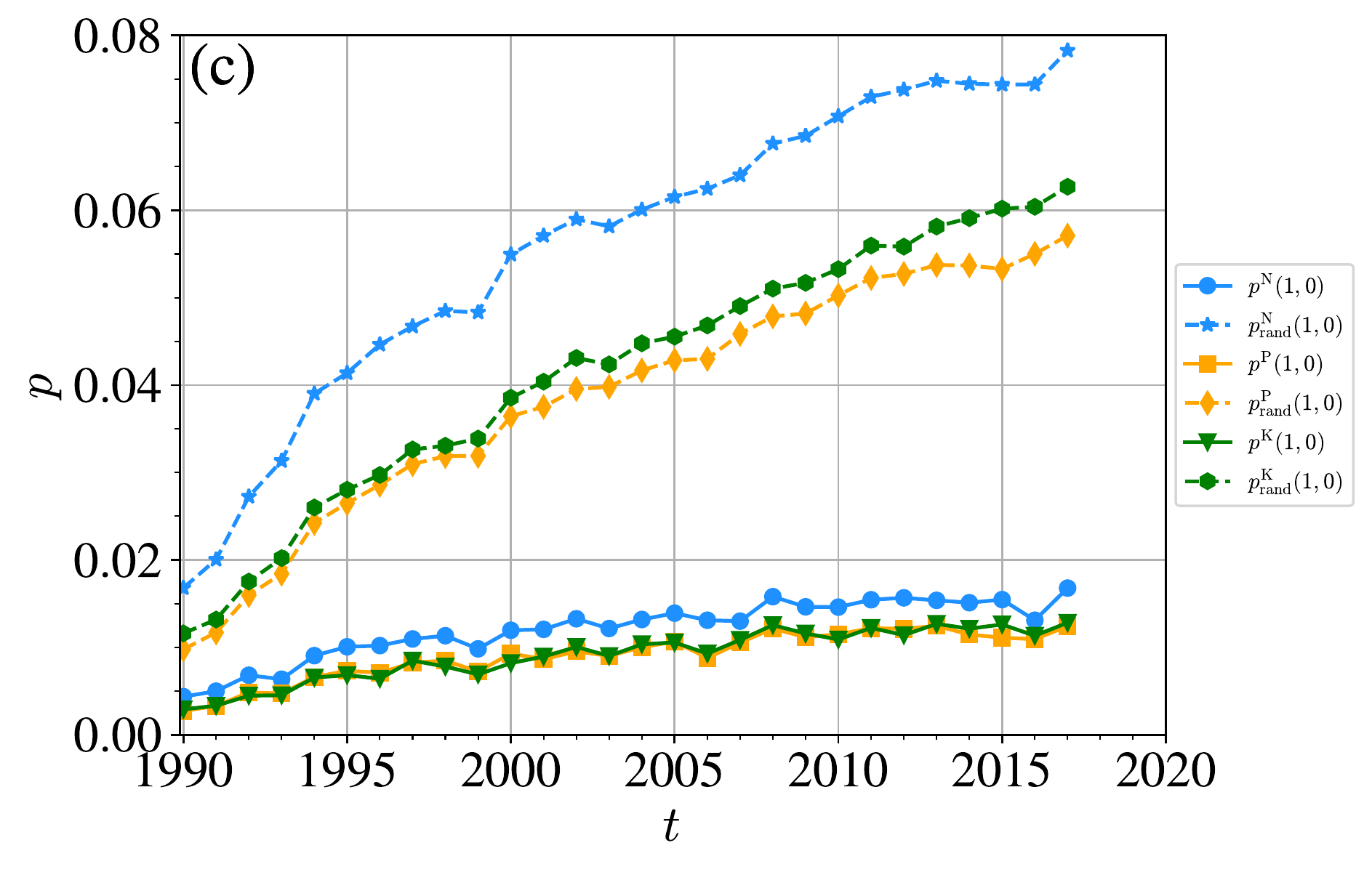}
    \includegraphics[width=0.49\linewidth]{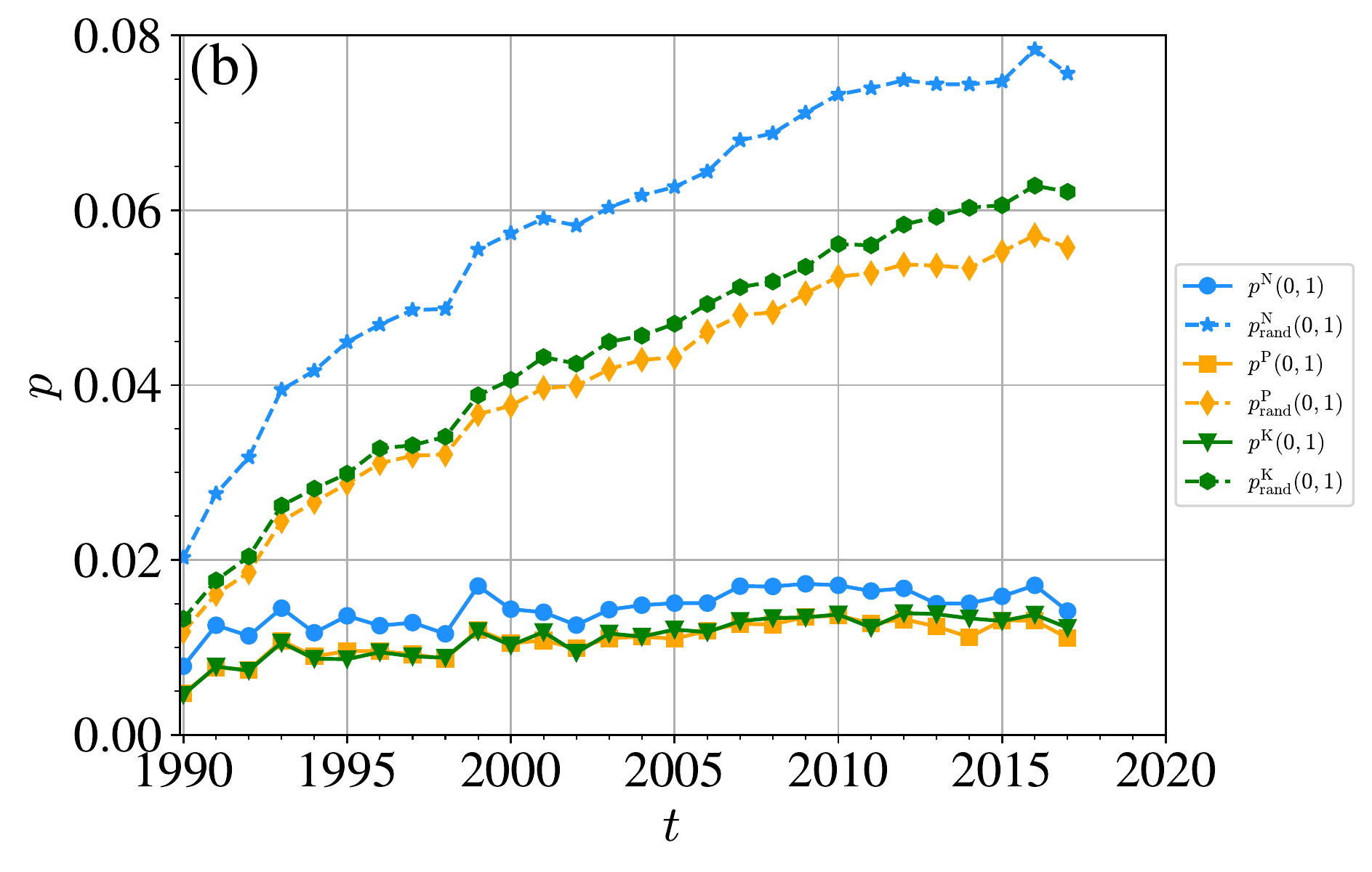}
    \includegraphics[width=0.49\linewidth]{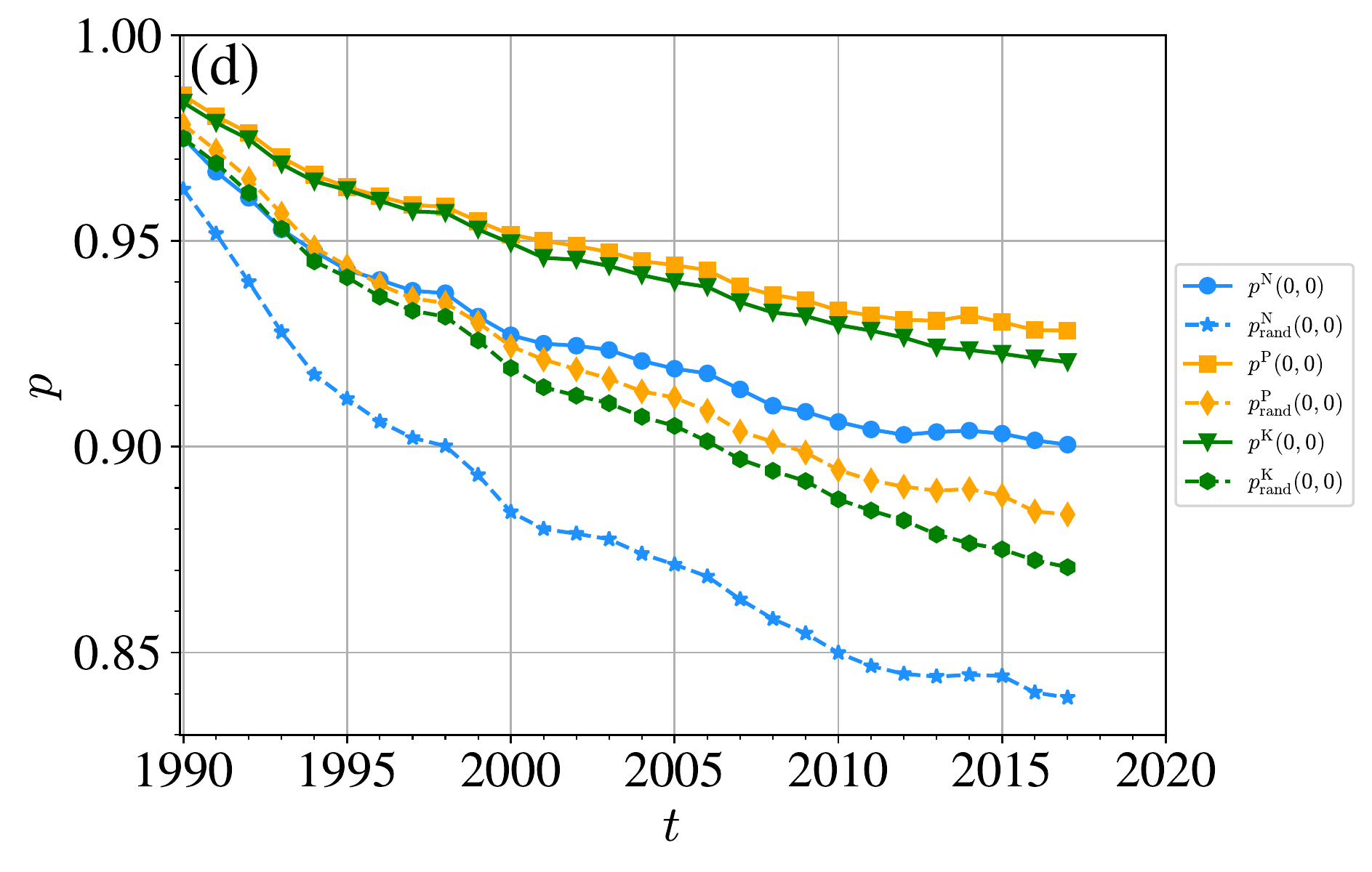}
    \caption{Evolution of the joint probability $p$ of links between two successive networks of the international N, P and K trade from 1990 to 2018. $p_{\mathrm{rand}}$ is the corresponding probability in an independent random network. $p(t, t+1)$ is the joint probability of $\mathscr{G}(t)$ and $\mathscr{G}(t+1)$, corresponding to the point $t$.
    }
    \label{Fig:FertilizerNet:Joint:Probablity}
\end{figure*}

Then we can calculate the the random joint probability $p_{\mathrm{rand}}(x_t, x_{t+1})$ of the two variables $x_t\in\{0,1\}$ and $x_{t+1}\in\{0,1\}$. The random joint probability $p_{\mathrm{rand}}(1,1)$ that a link $e_{ij}$ exists in both networks $\mathscr{G}^{'}(t)$ and $\mathscr{G}^{'}(t+1)$ is given by
\begin{equation}
  p_{\mathrm{rand}}(1,1) = p_{\mathscr{G}^{'}(t)}(1)  p_{\mathscr{G}^{'}(t+1)}(1) = \frac{N_{\mathscr{E}(t)}N_{\mathscr{E}(t+1)}}{N^2(N-1)}. 
\end{equation}
And the random joint probability $p_{\mathrm{rand}}(1, 0)$ that a link $e_{ij}$ exists in $\mathscr{G}^{'}(t)$ but not in $\mathscr{G}^{'}(t+1)$ is
\begin{equation}
  p_{\mathrm{rand}}(1,0) = p_{\mathscr{G}^{'}(t)}(1)  p_{\mathscr{G}^{'}(t+1)}(0) = \frac{N_{\mathscr{E}(t)}(1-N_{\mathscr{E}(t+1)})}{N^2(N-1)},
\end{equation}
while the joint probability $p_{\mathrm{rand}}(0, 1)$ that a link $e_{ij}$ does not exist in $\mathscr{G}^{'}(t)$ but exists in $\mathscr{G}^{'}(t+1)$ is given by
\begin{equation}
  p_{\mathrm{rand}}(0,1) = p_{\mathscr{G}^{'}(t)}(0) p_{\mathscr{G}^{'}(t+1)}(1) = \frac{(1 - N_{\mathscr{E}(t)}) N_{\mathscr{E}(t+1)}}{N^2(N-1)}. 
\end{equation}
We also have the joint probability $p_{\mathrm{rand}}(0, 0)$ that a link $e_{ij}$ does not exist in neither networks $\mathscr{G}^{'}(t)$ and $\mathscr{G}^{'}(t+1)$,
\begin{equation}
\begin{aligned}
  p_{\mathrm{rand}}(0,0) &= p_{\mathscr{G}^{'}(t)}(0)  p_{\mathscr{G}^{'}(t+1)}(0)  \\ &= 
  \frac{(1 - N_{\mathscr{E}(t)}) (1 - N_{\mathscr{E}(t+1)})}{N^2(N-1)}.
\end{aligned}
\end{equation}

Fig.~\ref{Fig:FertilizerNet:Joint:Probablity} describes the evolution of the four kinds of joint probability and corresponding four kinds of random joint probability of links between two successive networks of the three fertilizer nutrients from 1990 to 2018. 
In Fig.~\ref{Fig:FertilizerNet:Joint:Probablity}(a), we compare the joint probability $p(1,1)$ in the actual N, P, and K trade networks with the random joint probability $p_{\mathrm{rand}}(1,1)$ in the corresponding random networks. The former shows that the empirical joint probabilities are significantly higher than the random ones, that is, the probability that the links existing in $\mathscr{G}(t)$ remain in $\mathscr{G}(t+1)$ is much higher than the random case. This indicates that there is a strong inheritance relationship between the two continuous networks of actual trade. This effect becomes stronger along time. At the same time, we can find that in each year $p^{\mathrm{N}}(1,1)$ has a significantly larger value in the comparison of the three nutrients, which is because the density of the N trade network is much higher. Through the comparison of the joint probability $p(1,0)$ and the corresponding random joint probability $p_{\mathrm{rand}}(1,0)$ in Fig.~\ref{Fig:FertilizerNet:Joint:Probablity}(b), we can find that the probability that a link that exists in $\mathscr{G}(t)$ does not exist in $\mathscr{G}(t+1)$ is much lower than in the random case. This is as same as the result from the comparison of the joint probability $p(0,1)$ and the random joint probability $p_{\mathrm{rand}}(0,1)$ in Fig.~\ref{Fig:FertilizerNet:Joint:Probablity}(c). In Fig.~\ref{Fig:FertilizerNet:Joint:Probablity}(d)
, the joint probability $p(0,0)$ is much lower than the random joint probability $p_{\mathrm{rand}}(0,0)$. This implies that, in the actual network, the links that do not exist in $\mathscr{G}(t)$ are more difficult to appear in $\mathscr{G}(t+1)$.

By calculating the difference between the actual joint probability and the random joint probability, the four practical propensities can be obtained to describe the transfer situation between two network pairs after excluding the effect of random distribution. Here we name it as structural inheritance and use $r([1,0],[1,0])$ as marker. For the structural inheritance about the link $e_{ij}$ exists in both networks $\mathscr{G}(t)$ and $\mathscr{G}(t+1)$, we have
\begin{equation}
  r(1,1) = p(1,1) - p_{\mathrm{rand}}(1,1) =  \frac{N_{{\mathscr{E}}(t)\cap {\mathscr{E}}(t+1)}}{N(N-1)} - \frac{N_{\mathscr{E}(t)} N_{\mathscr{E}(t+1)}}{N^2(N-1)}.
\label{Eq:r11}
\end{equation}
For the structural inheritance about the link $e_{ij}$ exists in $\mathscr{G}(t)$ but not in $\mathscr{G}(t+1)$, we have
\begin{equation}
\begin{aligned}
  r(1,0) &= p(1,0) - p_{\mathrm{rand}}(1,0) \\&=   \frac{N_{\mathscr{E}(t)}-N_{{\mathscr{E}}(t)\cap {\mathscr{E}}(t+1)}}{N(N-1)} - \frac{(N_{\mathscr{E}(t)} (1-N_{\mathscr{E}(t+1)})}{N^2(N-1)}.
  \label{Eq:r10}
\end{aligned}
\end{equation}
For the structural inheritance about the link $e_{ij}$ does not exist in $\mathscr{G}(t)$ but exists in $\mathscr{G}(t+1)$, we have
\begin{equation}
\begin{aligned}
  r(0,1) &= p(0,1) - p_{\mathrm{rand}}(0,1) \\&=  \frac{N_{\mathscr{E}(t+1)}-N_{{\mathscr{E}}(t)\cap {\mathscr{E}}(t+1)}}{N(N-1)} - \frac{(1 - N_{\mathscr{E}(t)}) N_{\mathscr{E}(t+1)}}{N^2(N-1)}.
  \label{Eq:r01}
\end{aligned}
\end{equation}
For the structural inheritance about link $e_{ij}$ does not exist in neither networks $\mathscr{G}(t)$ and $\mathscr{G}(t+1)$, we have
\begin{equation}
\begin{aligned}
  r(0,0) &= p(0,0) - p_{\mathrm{rand}}(0,0) \\&=  1-\frac{N_{\mathscr{E}(t)}+N_{\mathscr{E}(t+1)}-N_{{\mathscr{E}}(t)\cap {\mathscr{E}}(t+1)}}{N(N-1)}  \\ &\ \ \ \   - \frac{(1 - N_{\mathscr{E}(t)}) (1 - N_{\mathscr{E}(t+1)})}{N^2(N-1)}
  \label{Eq:r00}
\end{aligned}  
\end{equation}

Eqs.~(\ref{Eq:r11}-\ref{Eq:r00}) are the direct expressions for transfer situation. These four expressions can be linked mathematically. For $r(1,1)$, the format can be decomposed as
\begin{equation}
\begin{aligned}
  r(1,1) &= p(1,1) - p_{\mathrm{rand}}(1,1)    
  \\&= p_{\mathscr{G}(t)}(1) - p(1,0) -  \left[ p_{\mathscr{G}^{'}(t)}(1) - p_{\mathrm{rand}}(1,0)\right]
\end{aligned}
\end{equation}
Since $\mathscr{G}(t)$ and $\mathscr{G}^{'}(t)$ have exactly the same network density,
\begin{equation}
 p_{\mathscr{G}(t)}(1)=p_{\mathscr{G}^{'}(t)}(1),
\label{Eq:pgi=pgii}
\end{equation}
we can get the substitution about
\begin{equation}
\begin{aligned}
  p_{\mathrm{rand}}(1,0)&=p_{\mathscr{G}^{'}(t)}(1) (1-p_{\mathscr{G}^{'}(t+1)}(1))
  \\&=p_{\mathscr{G}^{'}(t)}(1) - p(1,0) 
  \\&= p_{\mathscr{G}(t)}(1) - p(1,0),
\end{aligned}
\end{equation}
and
\begin{equation}
  p(1,0)=p_{\mathscr{G}(t)}(1) -p_{\mathrm{rand}}(1,0)
  =p_{\mathscr{G}^{'}(t)}(1) - p_{\mathrm{rand}}(1,0).
\end{equation}
We have
\begin{equation}
  r(1,1)=p_{\mathrm{rand}}(1,0) - p(1,0)
  =-r(1,0).
\label{Eq:r11:r10}
\end{equation}
Similarly, $r(1,1)$ can also be decomposed into
\begin{equation}
\begin{aligned}
  r(1,1) &= p(1,1) - p_{\mathrm{rand}}(1,1)
  \\&= p_{\mathscr{G}(t+1)}(1) - p(0,1) -  \left[ p_{\mathscr{G}^{'}(t+1)}(1) - p_{\mathrm{rand}}(0,1)\right].
  \end{aligned}
\end{equation}
Because $\mathscr{G}(t+1)$ and $\mathscr{G}^{'}(t+1)$ also have exactly the same network density,
\begin{equation}
 p_{\mathscr{G}(t+1)}(1)=p_{\mathscr{G}^{'}(t+1)}(1),
\end{equation}
we have
\begin{equation}
  r(1,1)=p_{\mathrm{rand}}(0,1) - p(0,1)
    =-r(0,1).
\label{Eq:r11:r01}
\end{equation}
For $r(0,0)$, the format is expanded as
\begin{equation}
\begin{aligned}
  r(0,0) &= p(0,0) - p_{\mathrm{rand}}(0,0)
  \\&= p_{\mathscr{G}(t)}(0) - p(0,1) -  \left[ p_{\mathscr{G}^{'}(t)}(0) - p_{\mathrm{rand}}(0,1)\right],
\end{aligned}
\end{equation}
According to Eq.~(\ref{Eq:pgi=pgii}), we can get
\begin{equation}
  r(0,0)=p_{\mathrm{rand}}(0,1) - p(0,1)
  =-r(0,1)
  =r(1,1).
\label{Eq:r00:r11}
\end{equation}
Combining Eqs.~(\ref{Eq:r11:r10}), (\ref{Eq:r11:r01}), and (\ref{Eq:r00:r11}), we finally get the mathematical relationship of the four expressions:
\begin{equation}
\begin{aligned}
  r(0,0)&=r(1,1)=-r(1,0)=-r(0,1) \\&= p(1, 1) - p_{\mathscr{G}(t)}(1)   p_{\mathscr{G}(t+1)}(1).
\label{Eq:rho}
\end{aligned}
\end{equation}

It has been proved that the four link transition probabilities between network pairs have a unified expression after excluding the influence of random distribution. On this basis, in order to make the $r$ values comparable between different network pairs, we use the maximum similarity state of the network pair as $p_{\max}$ for normalization:
\begin{equation}
  r = \frac{p(1, 1) - p_{\mathrm{rand}}(1, 1)}{p_{\max}(1, 1) - p_{\mathrm{rand}}(1, 1)}.
\label{Eq:rho_s}
\end{equation}
In the end, by calculating the difference between the actual joint probability and the random joint probability, we obtain the unified normalized structural inheritance $r$ of the four tendencies after excluding the influence of random distribution. For two successive networks $\mathscr{G}(t)$ and $\mathscr{G}(t+1)$, we use $r(t, t+1)$ to express the structural inheritance between the two networks.

\begin{figure*}[!ht]
    \centering
    \includegraphics[width=0.49\linewidth]{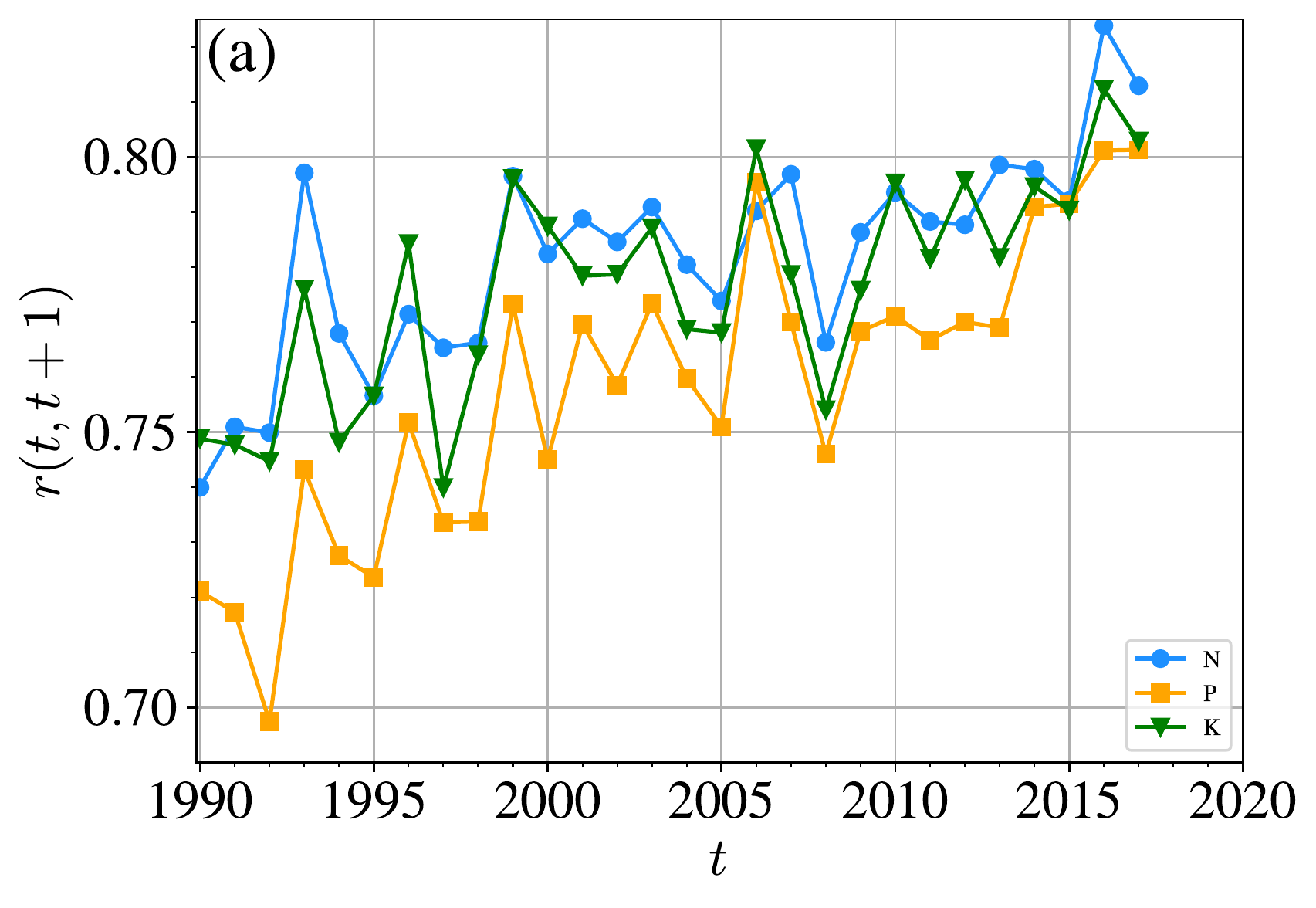}        \includegraphics[width=0.49\linewidth]{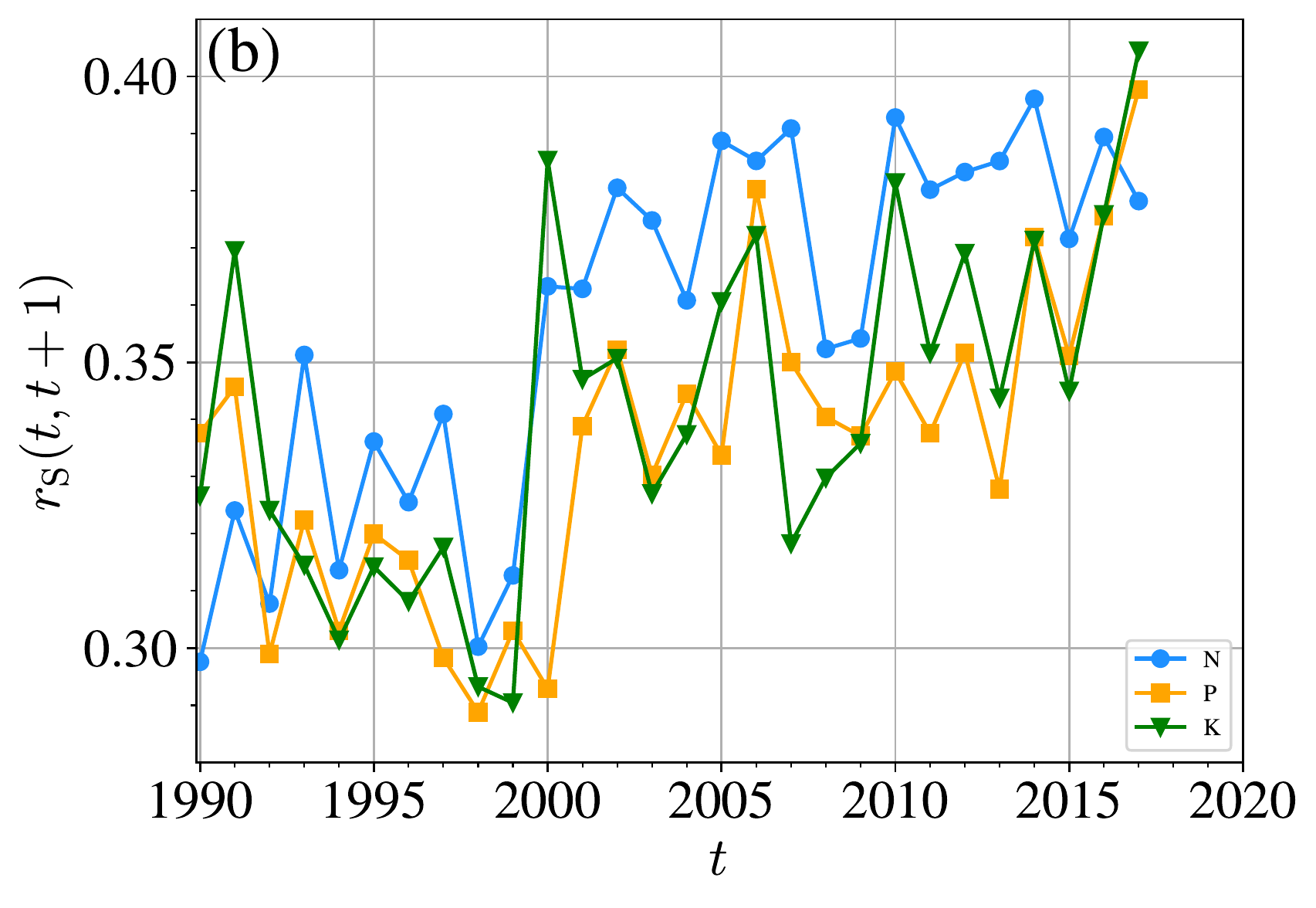}   
    \includegraphics[width=0.49\linewidth]{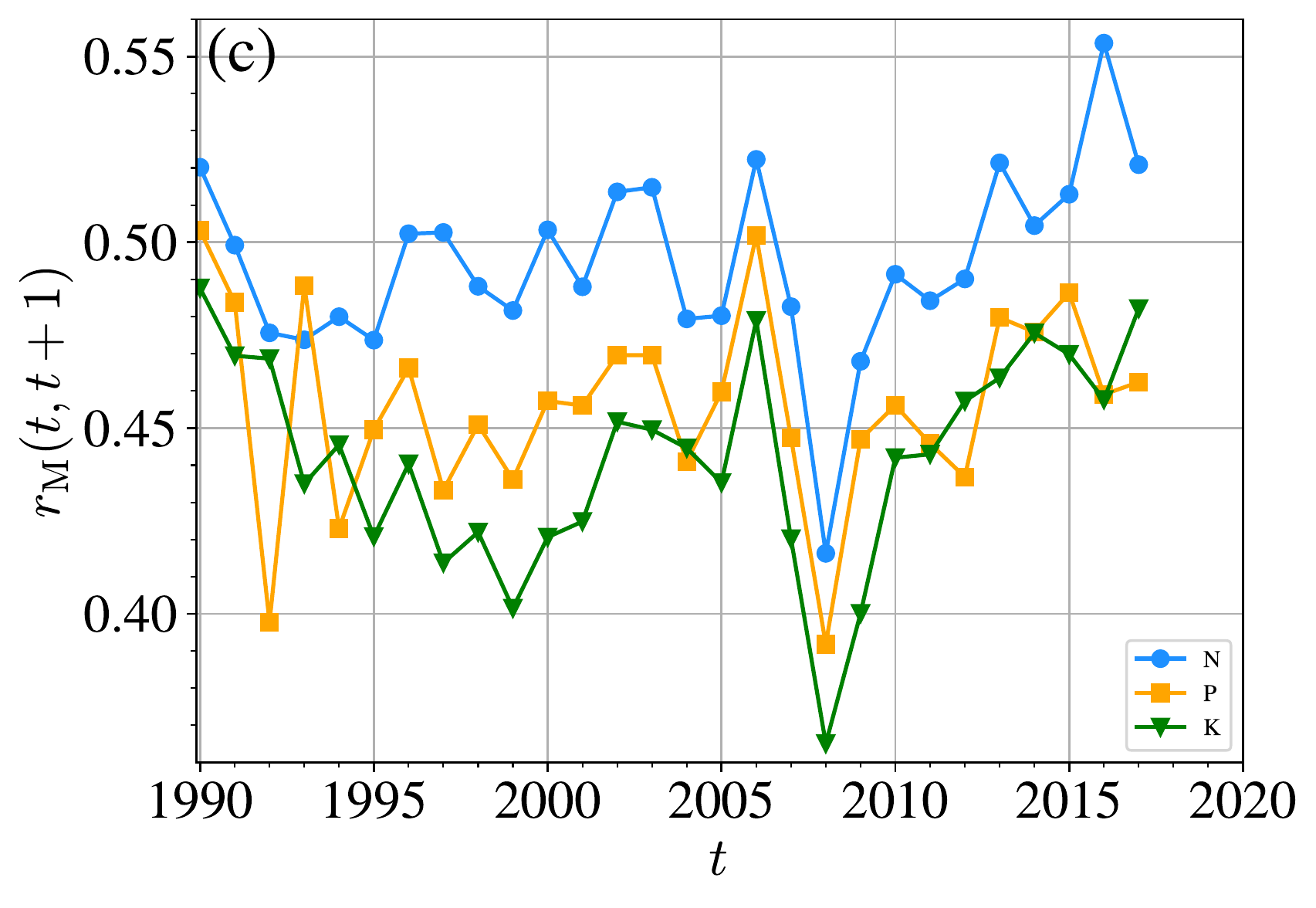}   
    \includegraphics[width=0.49\linewidth]{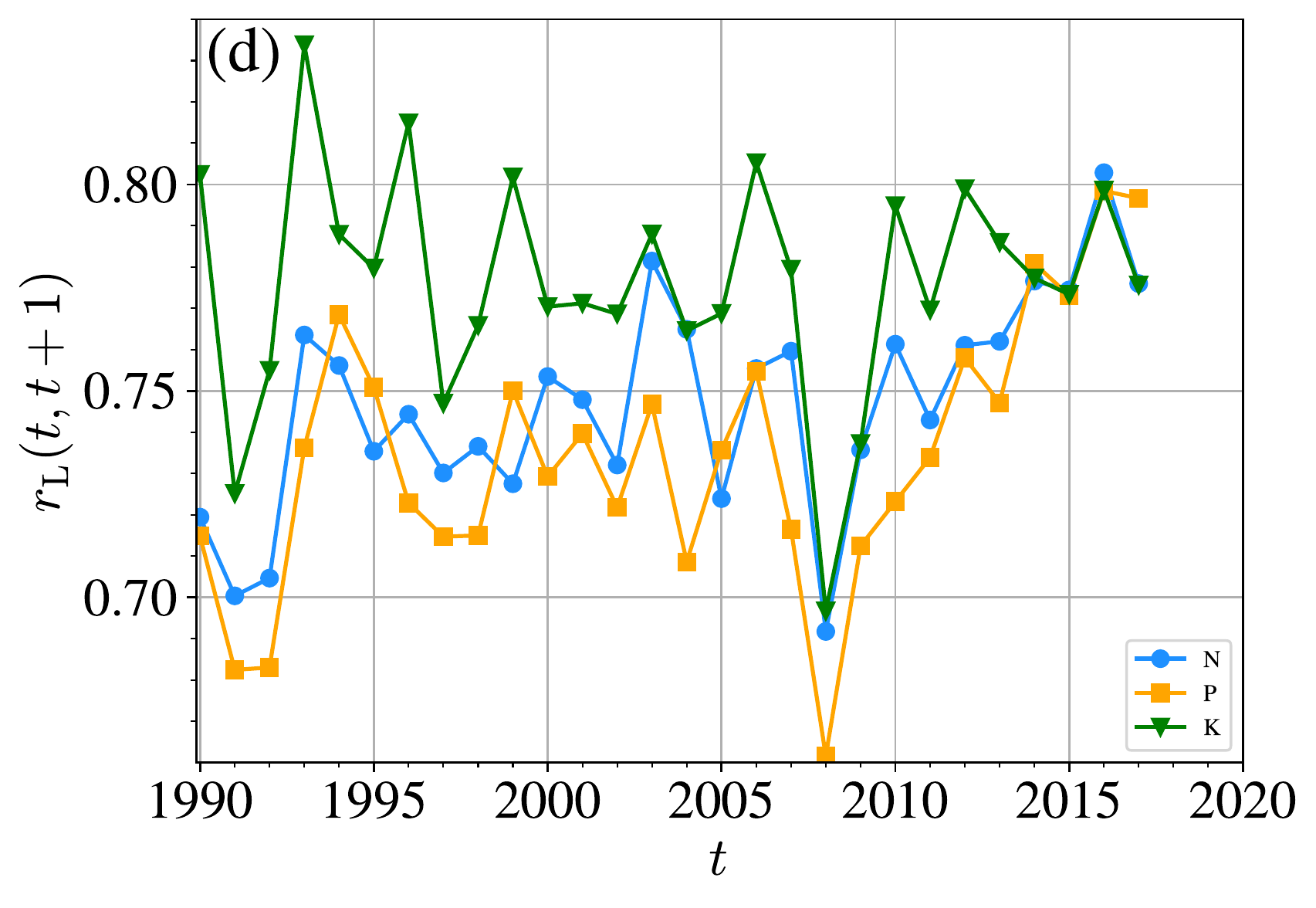}   
    \caption{Evolution of the normalized structural inheritance $r(t,t+1)$ between two successive trade networks of the international N, P and K trade from 1990 to 2018. (a) Overall network containing all links. (b) Sub-network containing small links with the weights less than the 20\% percentile. (c) Sub-network containing medium links with the weights between the 40\% and 60\% percentiles. (d) Sub-network containing large links with the weights greater than the 80\% percentile.}
    \label{Fig:FertilizerNet:rho}
\end{figure*}

We plot $r(t, t+1)$ against $t$ of the three nutrient trade networks in Fig.~\ref{Fig:FertilizerNet:rho}(a). Obviously, the $r(t, t+1)$ functions of the three networks are much greater than 0, which means that the structure in $\mathscr{G}(t)$ has a high probability of being inherited to $\mathscr{G}(t+1)$. Each $r(t, t+1)$ curve has an upward trend with time $t$, indicating that the stability of the international fertilizer trade networks as a whole is increasing. While the calculation of the structural inheritance excludes the impact of changes in network density, the value of $r(1991, 1992)$ didn't drop significantly caused by the disintegration of the Soviet Union. Correspondingly, the 2008 economic crisis has a huge systematic impact on the trade system. From 2008 to 2009, many trades broke down, so the $r(2008, 2009)$ values of the three nutrient trade networks all decreased significantly. If a large-scale shock at time $t$ happens to break the links in the network and reduce $r(t, t+1)$, the network structure retained after the shock will be more stable, causing $r(t+1, t+2)$ to have higher values. For example, $r(2006, 2007)$ reached a short-term peak after a significant decline in $r(2005, 2006)$. After the drop in $r(2008, 2009)$, $r(2009,2010)$ still showed a significant rebound despite the lingering effects of the economic crisis. And the fall of $r(2015,2016)$ under the European debt crisis also contributed to the highest value of $r(2016,2017)$. The links that have been maintained after fluctuations have more retention ability, creating an upward trend of $r(t, t+1)$ over time.

Horizontally comparing the structural inheritance $r(t, t+1)$ of the three fertilizers, we find
\begin{equation}
    r^{\mathrm{N}}(t, t+1)\approx r^{\mathrm{K}}(t, t+1) > r^{\mathrm{P}}(t, t+1).
\end{equation}
Although the network density of the K trade network is much lower than that of the N trade network, their structural inheritances $r(t, t+1)$ are close, implying that the ratio of stable structure in the N network and K network are similar. The $r^{\mathrm{P}}(t, t+1)$ in the P trade is significantly low in the early stage, and gradually approach the other two in the later stage. The time trends of the structural inheritance of the three nutrients remain basically the same. An interesting point is that both $r^{\mathrm{P}}(2007, 2008)$ and $r^{\mathrm{K}}(2007, 2008)$ started to decline due to the economic crisis, while $r^{\mathrm{N}}(2007, 2008)$ continued to increase to reach a short-term maximum.


While Fig.~\ref{Fig:FertilizerNet:rho}(a) reflects the evolution of structural inheritance of the network as a whole, we subdivide the structural inheritance of different parts of the network according to the weights of the links for in-depth analysis. Fig.~\ref{Fig:FertilizerNet:rho}(b) describes the structural inheritance $r_{\mathrm{S}}(t, t+1)$ of the sub-network containing links with the weights below the 20\% percentile. Fig.~\ref{Fig:FertilizerNet:rho}(c) describes the structural inheritance $r_{\mathrm{M}}(t, t+1)$ of the sub-network containing links with the weights above the 40\% percentile and below the 60\% percentile. Fig.~\ref{Fig:FertilizerNet:rho}(d) describes the structural inheritance $r_{\mathrm{L}}(t, t+1)$ of the sub-network containing links with the weights above the 80\% percentile, which play the most important role in guaranteeing trade system functionality. Comparing Fig.~\ref{Fig:FertilizerNet:rho}(b-d), we observe that the sub-network with large-weight links is more likely to retain its own structure in $\mathscr{G}(t+1)$. This means that more important parts of the network have greater stability, which is a very helpful property for the long-term operation of the system. On the other hand, it is surprising that the structural inheritance of some sub-networks, such as $r^{\mathrm{K}}_{\mathrm{L}}(t, t+1)$, does not show a trend over time according to the ADF test, which is different from the results of the overall network. This shows that the evolution of the stability of different parts of the network is not synchronized with the evolution of the global stability.

Comparing the evolution of $r_{\mathrm{S,M,L}}(t, t+1)$ for N, P and K trade, $r^{\mathrm{K}}_{\mathrm{L}}(t, t+1)$ is the largest. We find that
\begin{equation}
    r^{\mathrm{K}}_{\mathrm{L}}(t, t+1) > r^{\mathrm{N}}_{\mathrm{L}}(t, t+1) \approx r^{\mathrm{P}}_{\mathrm{L}}(t, t+1),
\end{equation}
which means that the large-weight trade sub-network of K has relatively higher stability. For the middle-weight part, the $r^{\mathrm{N}}_{\mathrm{M}}(t, t+1)$ is higher and we have
\begin{equation}
    r^{\mathrm{N}}_{\mathrm{M}}(t, t+1) > r^{\mathrm{K}}_{\mathrm{M}}(t, t+1) \approx r^{\mathrm{P}}_{\mathrm{M}}(t, t+1),
\end{equation}
indicating that the stability of the mediate-weight trade sub-network of N is the main reason why its overall network has a high structural inheritance $r(t, t+1)$. In the evolution of the structural inheritance $r_{\mathrm{S}}(t, t+1)$ of small-weight part, it can be find the differences between the three fertilizers are not significant, that is,
\begin{equation}
    r^{\mathrm{N}}_{\mathrm{S}}(t, t+1) \approx r^{\mathrm{K}}_{\mathrm{S}}(t, t+1) \approx r^{\mathrm{P}}_{\mathrm{S}}(t, t+1),
\end{equation}
reflecting the commonality of instability in small-scale trade sub-networks.

In the evolution of $r_{\mathrm{S}}(t, t+1)$ in Fig.~\ref{Fig:FertilizerNet:rho}(b), the largest increase of N and K nutrient trade occurs between $r_{\mathrm{S}}^{\mathrm{N,K}}(1999, 2000)$ and $r_{\mathrm{S}}^{\mathrm{N,K}}(2000, 2001)$. However, the structural inheritance $r^{\mathrm{N,K}}(1999, 2000)$ of the overall network shows the opposite trend, implying that the structural inheritance of small-scale part and the structural inheritance of the whole system are out of synchronization in the early stage. But the trend of $r_{\mathrm{S}}(t, t+1)$ in the later stage is more similar to that of $r(t, t+1)$, which is an interesting point. As trading systems become more mature, the correlation between the overall network and the small-scale part becomes stronger.

From the evolution of $r_{\mathrm{M}}(t, t+1)$ in Fig.~\ref{Fig:FertilizerNet:rho}(c), we can find that $r_{\mathrm{M}}(t, t+1)$ is slightly higher than $r_{\mathrm{S}}(t, t+1)$, and much lower than the value of $r(t, t+1)$. Since the trade weight distribution has a long tail, there is actually a very big difference between large trade and medium trade. On the other hand, the values of $r^{\mathrm{N,P,K}}_{\mathrm{M}}(t, t+1)$ are relatively stable, except for the big drop between 2007 and 2009 to a minimum. Comparing Fig.~\ref{Fig:FertilizerNet:rho}(d), it can be found that the food crisis in 2007-2009 \cite{Gotz-Glauben-Brummer-2013-FoodPolicy} and the financial crisis in 2008-2010 \cite{Jiang-Zhou-Sornette-Woodard-Bastiaensen-Cauwels-2010-JEconBehavOrgan} has a very strong impact on the trade of medium and large weights in the fertilizer trading system, while the impact on the small-weight part is much lower.

The structural inheritance curves $r_{\mathrm{L}}(t, t+1)$ of the large-weight sub-networks in Fig.~\ref{Fig:FertilizerNet:rho}(d) have high values. This indicates the large-weight part which undertakes the main function of fertilizer trade system has the high stability. Its evolution process is the closest to the result of $r(t, t+1)$ in Fig.~\ref{Fig:FertilizerNet:rho}(a), which shows the large-weight link has a huge effect on the overall system. It is worth noting that $r^{\mathrm{N}}_{\mathrm{L}}(2007,2008)$ is the only increasing part in that year, which has mapped to $r^ {\mathrm{N}}(2007, 2008)$. The collapse of the Soviet Union had a great impact on the large-weight part from 1991 to 1992. It can be seen that $r^{\mathrm{N,P,K}}_{\mathrm{L}}(1991, 1992)$ were very low. But the impacts on the medium-weight and small-weight parts were unexpectedly small, implying that the impact of the collapse of the Soviet Union is mainly on the core of the system.

Summarizing Fig.~\ref{Fig:FertilizerNet:rho}, the structural inheritance we used can characterize the transfer process of $\mathscr{G}(t)$ to $\mathscr{G}(t+1)$ well. It can be used as an indicator to measure the structural inheritance in quantifying the temporal stability of networks. In the evolution of the three nutrient trade networks, this indicator is consistent with important historical events and reveals deep information. Next we continue to quantify the temporal stability of international fertilizer trade networks by using the other two methods for comparison.


\begin{figure*}[!ht]
    \centering
    \includegraphics[width=0.49\linewidth]{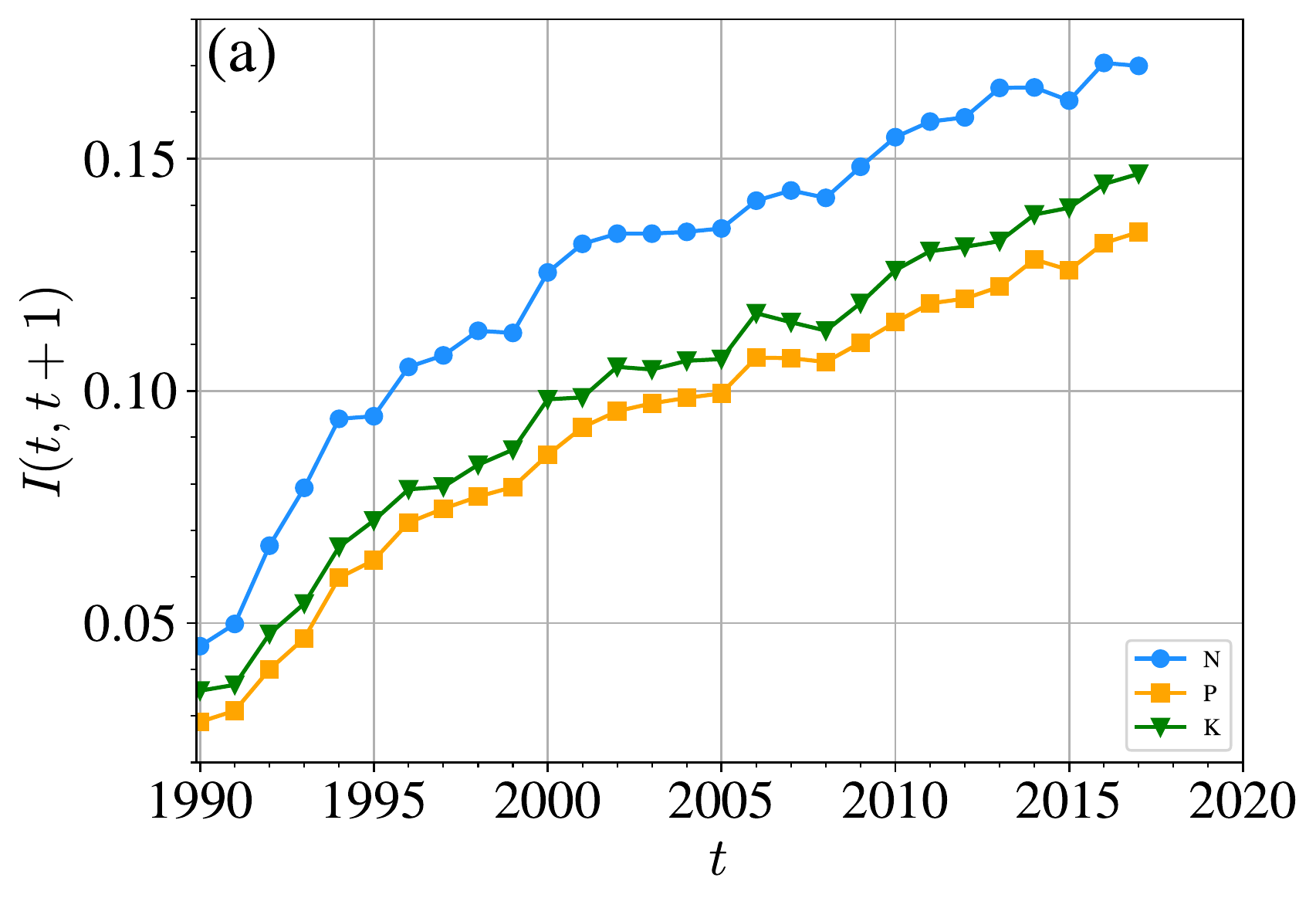}
    \includegraphics[width=0.49\linewidth]{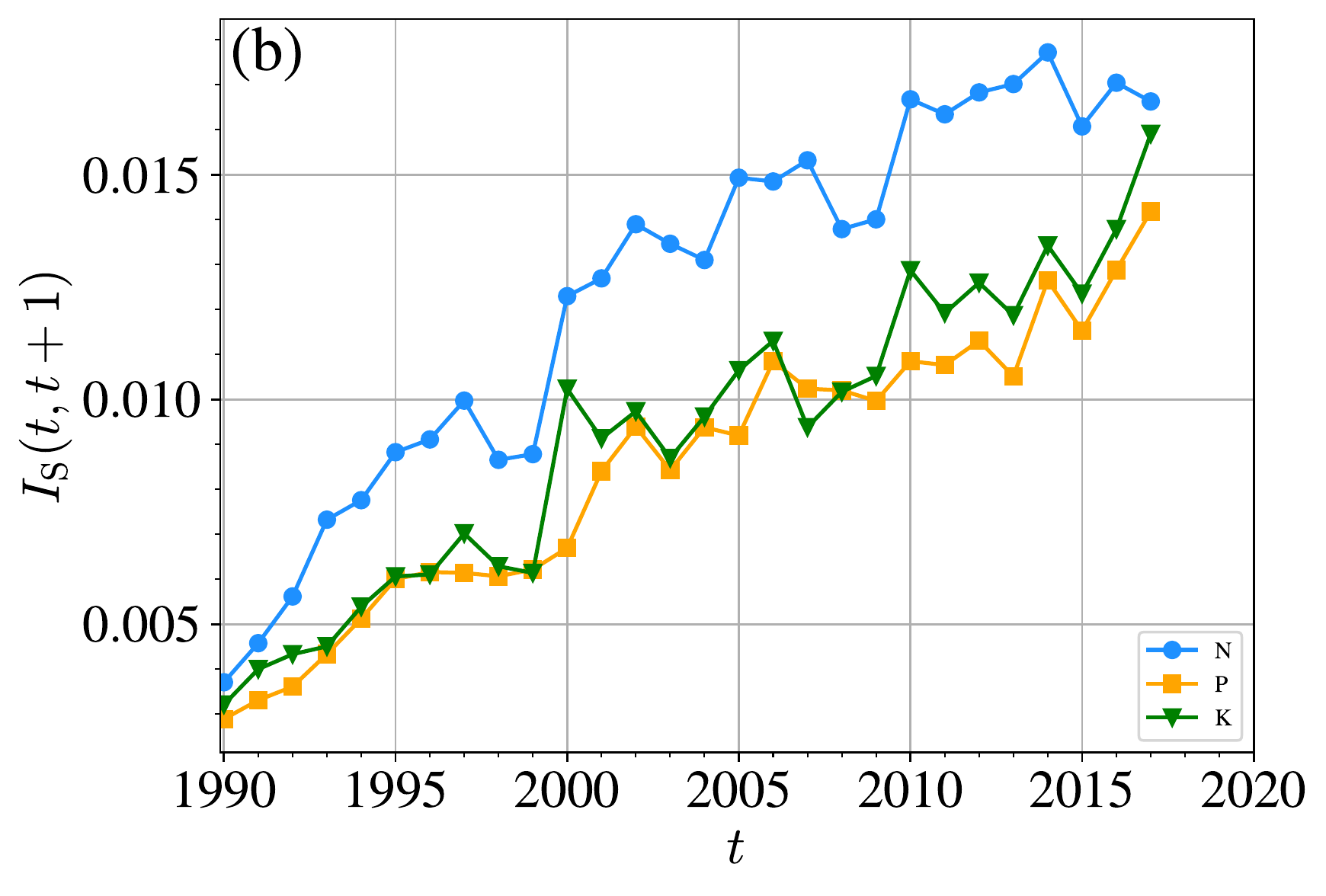}
    \includegraphics[width=0.49\linewidth]{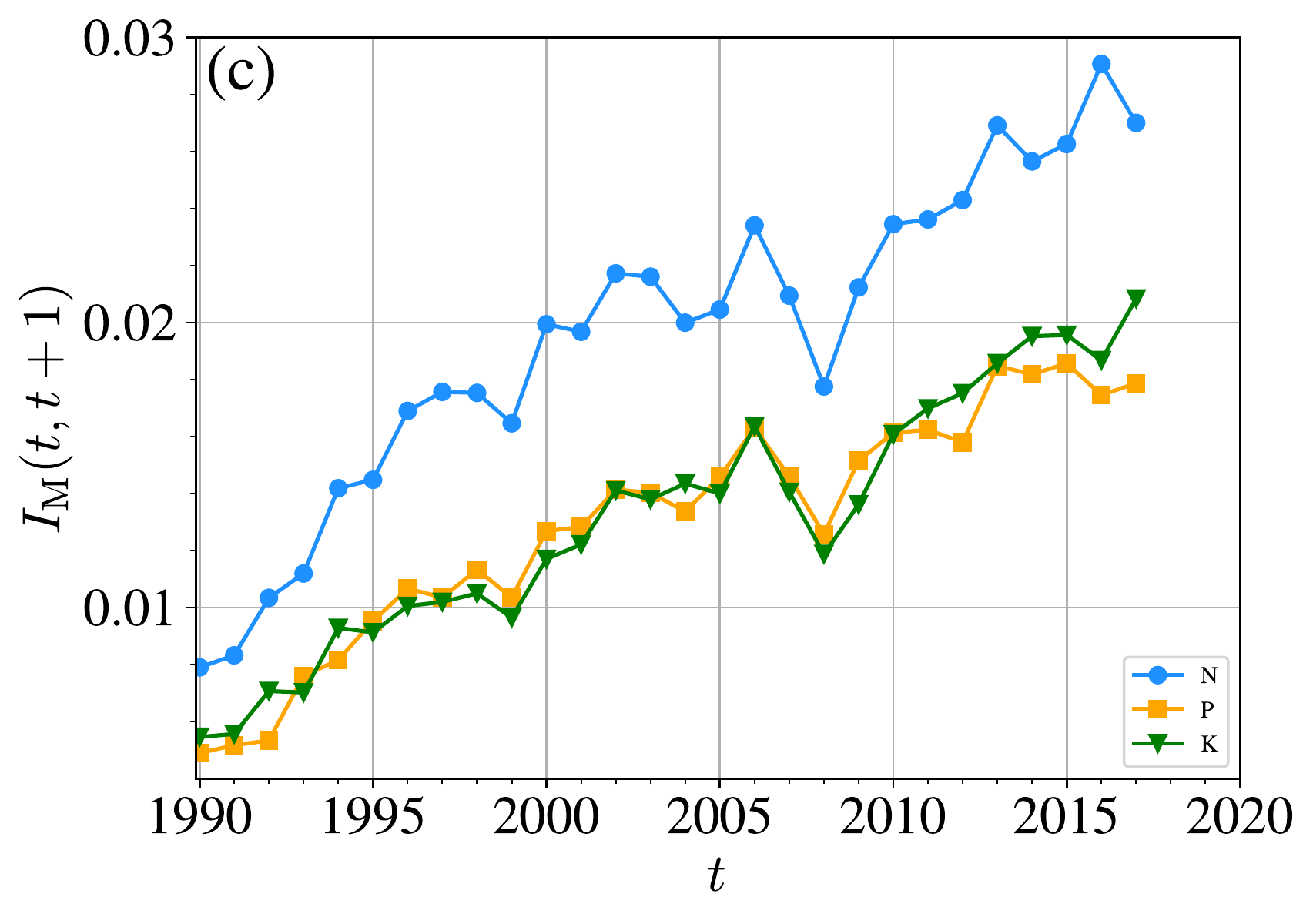}
    \includegraphics[width=0.49\linewidth]{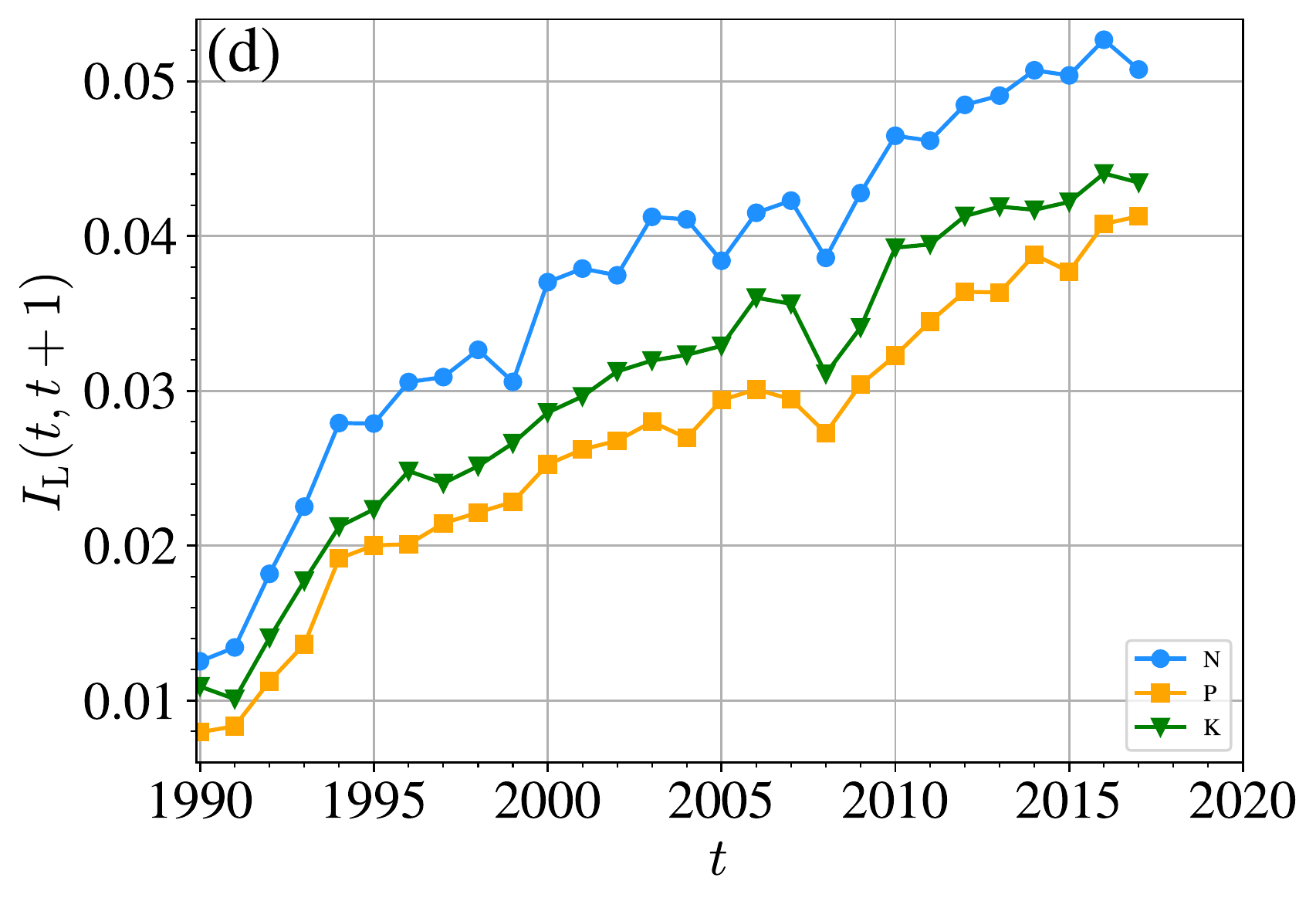}
    \caption{Evolution of the mutual information $I(t,t+1)$ between two successive networks of the international N, P and K trade from 1990 to 2018. (a) Overall network containing all links. (b) Sub-network containing small links with the weights less than the 20\% percentile. (c) Sub-network containing medium links with the weights between the 40\% and 60\% percentiles. (d) Sub-network containing large links with the weights greater than the 80\% percentile.}
    \label{Fig:FertilizerNet:I}
\end{figure*}

\subsection{Mutual information}

Another dimension of the stability of the network is to judge the correlation between network structures. Here we use mutual information to quantify the temporal stability of the international fertilizer trade networks \cite{Song-Tumminello-Zhou-Mantegna-2011-PhysRevE,Schieber-Carpi-Frery-Rosso-Pardalos-Ravetti-2016-PhysLettA}.


The mutual information of the random variables $x$ and $y$ is given by
\begin{equation}
  I(x,y)= \sum_{x} \sum_{y} p(x,y) \log \frac{p(x, y)}{p(x)p(y)}.
\end{equation}
During the transfer from $\mathscr{G}(t)$ to $\mathscr{G}(t+1)$, the mutual information between $\mathscr{G}(t)$ and $\mathscr{G}(t+1)$ is
\begin{equation}
  I(t,t+1)= \sum_{x_t=0}^{1} \sum_{x_{t+1}=0}^1 p(x_t,x_{t+1}) \log \frac{p(x_t, x_{t+1})}{p_{\mathscr{G}(t)}(x_t)p_{\mathscr{G}(t+1)}(x_{t+1})}.
\end{equation}
The calculation of $p(x_t,x_{t+1})$ is the same as in Eqs.~(\ref{Eq:p11}-\ref{Eq:p00}), while $p_{\mathscr{G}(t)(x_t)}$ and $p_{\mathscr{G}(t+1)}(x_{t+1})$ are calculated as in Eqs.~(\ref{Eq:p(1)}) and (\ref{Eq:p(0)}).

In Fig.~\ref{Fig:FertilizerNet:I}, we describe the evolution of the mutual information $I(t,t+1)$ between two successive networks of N, P, K trade from 1990 to 2018. Fig.~\ref{Fig:FertilizerNet:I}(a) describes the mutual information $I(t,t+1)$ of $\mathscr{G}(t)$ and $\mathscr{G}(t+1)$ containing all links, Fig.~\ref{Fig:FertilizerNet:I}(b) describes the mutual information $I_{\mathrm{S}}(t,t+1)$ of sub-network containing links with weights below the 20\% percentile, Fig.~\ref{Fig:FertilizerNet:I}(c) describes the mutual information $I_{\mathrm{M}}(t,t+1)$ of sub-network containing links with weights above the 40\% percentile and below the 60\% percentile, and Fig.~\ref{Fig:FertilizerNet:I}(d) describes the mutual information $I_{\mathrm{L}}(t,t+1)$ of sub-network containing links with weights above the 80\% percentile. 

For each case, the mutual information curve between $\mathscr{G}(t)$ and $\mathscr{G}(t+1)$ has a consistent increasing trend of with time. This is due to the fact that the trading system is becoming more and more prosperous with the process of globalization. It is the inevitable result of the increasing network density of the trading system. Therefore, we need to normalize the mutual information to observe the exact correlation between $\mathscr{G}(t)$ and $\mathscr{G}(t+1)$ to characterize the transfer process. Comparing the four plots, there are sharp drops in the mutual information in 2007 and 2008 for the large- and medium-weight sub-networks, showing that medium- and large-scale trades are significantly impacted by the soaring food price and the resulting food crisis, while small-scale trades receive a relatively low impact.

\begin{figure*}[htb]
    \centering
    \includegraphics[width=0.49\linewidth]{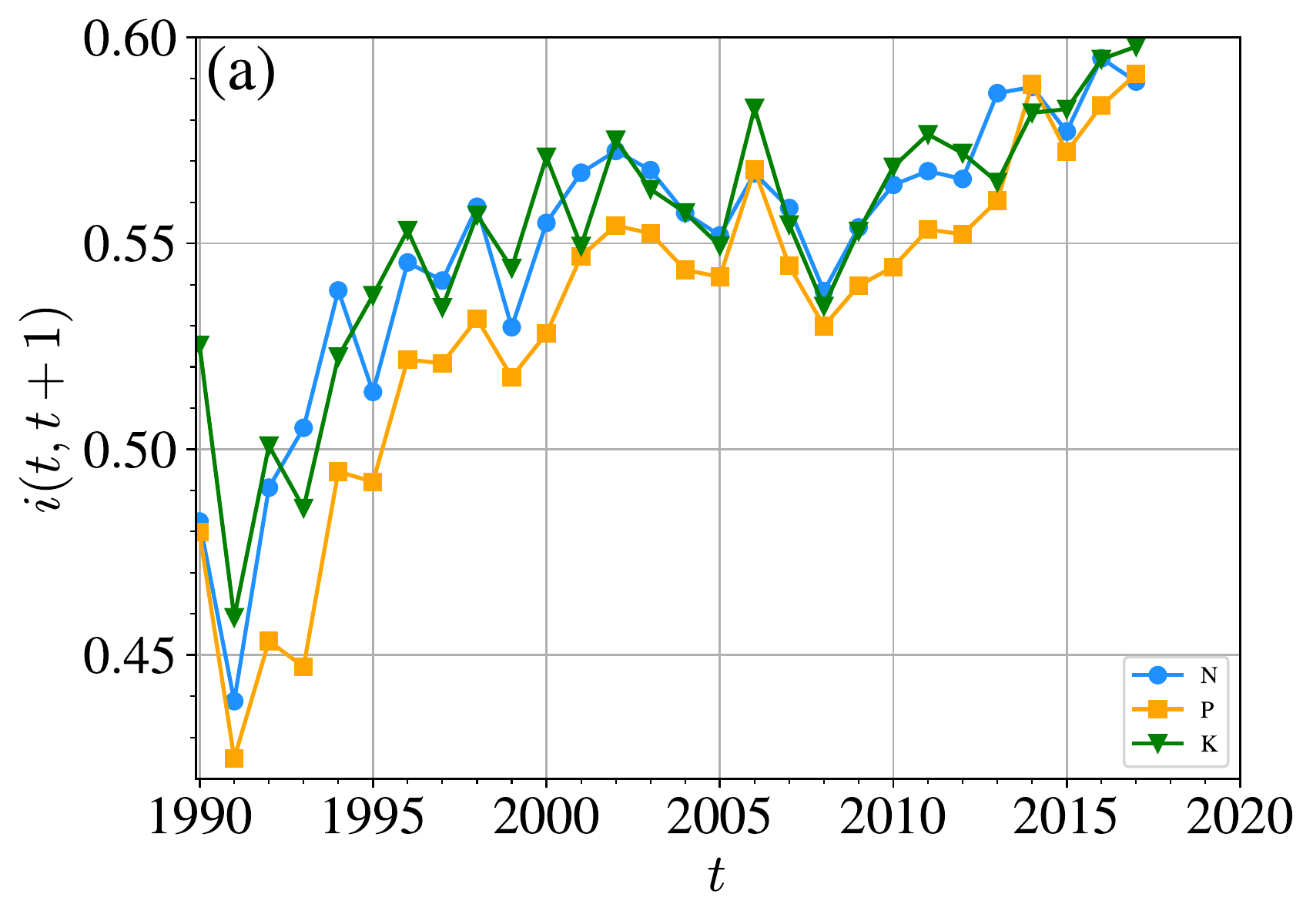}
    \includegraphics[width=0.49\linewidth]{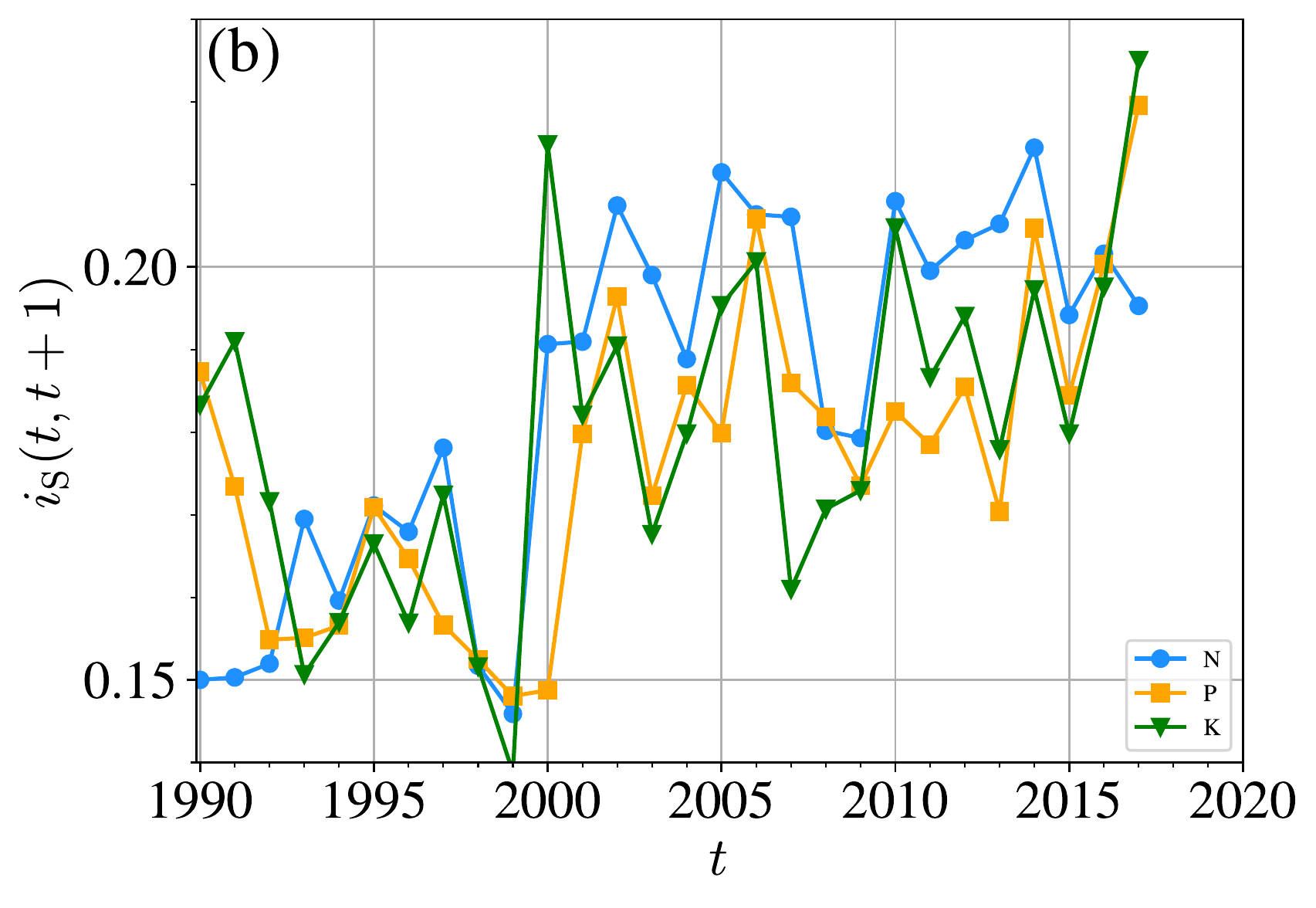}
    \includegraphics[width=0.49\linewidth]{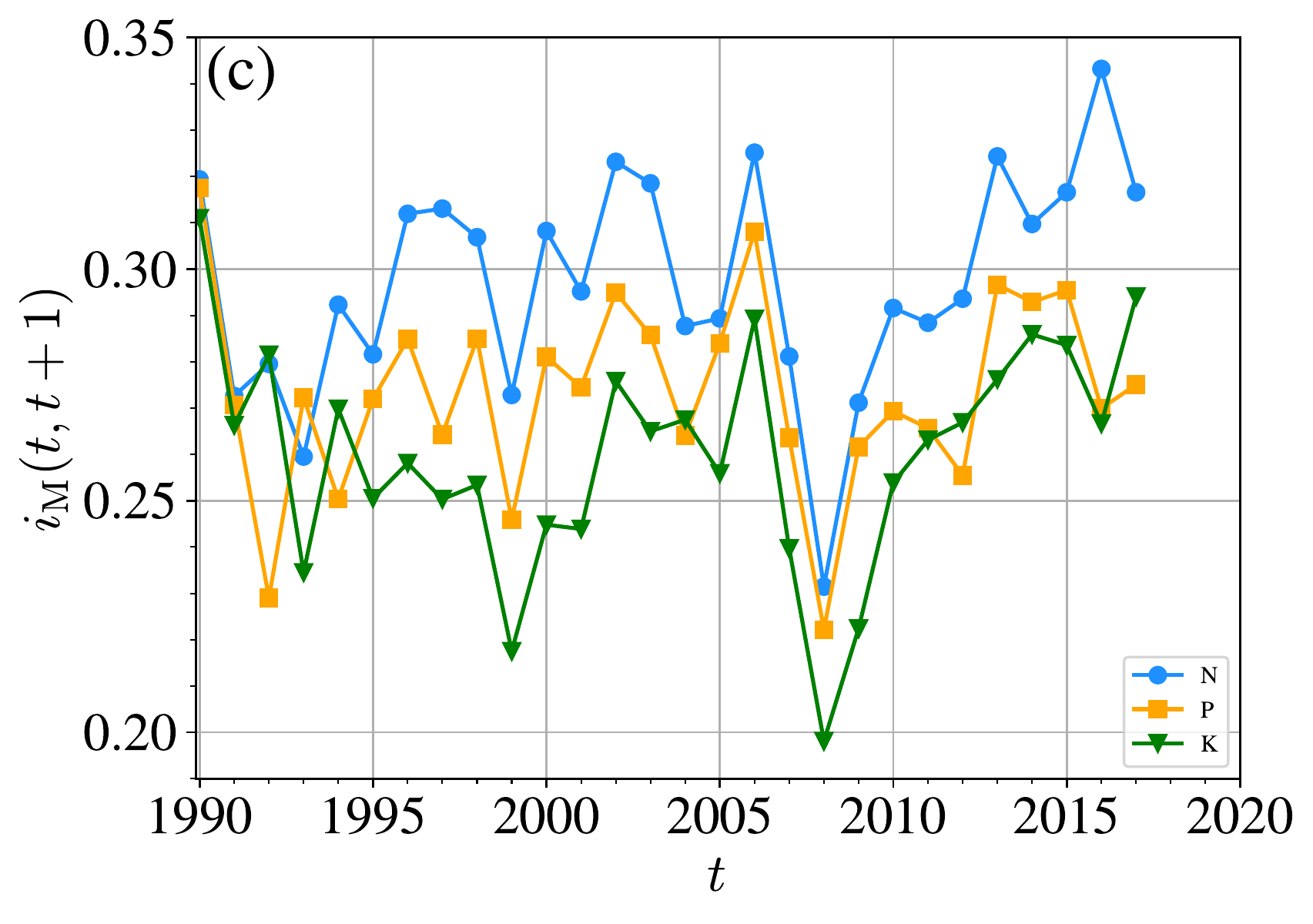}
    \includegraphics[width=0.49\linewidth]{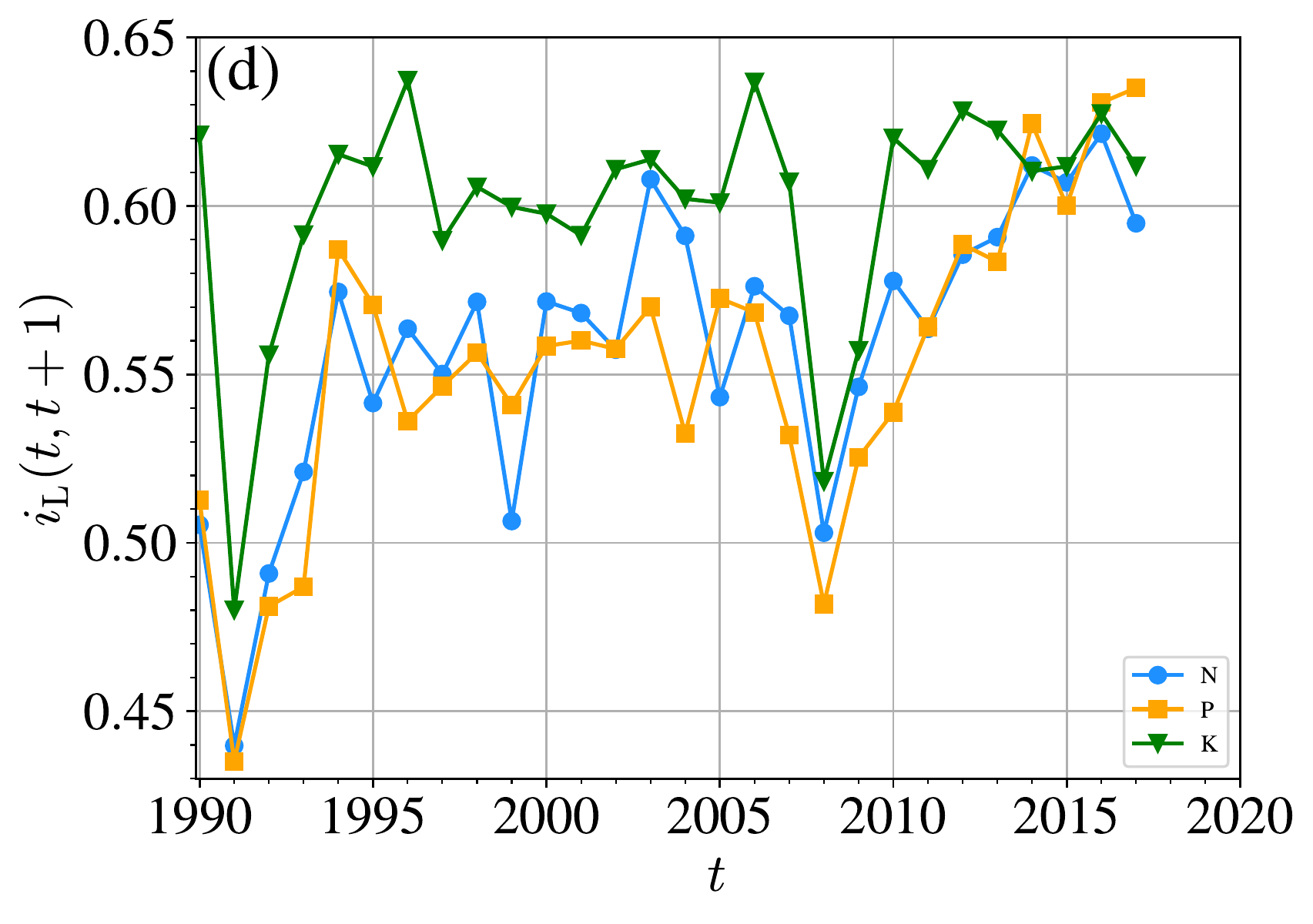}
    \caption{Evolution of the normalized the mutual information $i(t,t+1)$ between two successive networks of the international N, P and K trade from 1990 to 2018. (a) Overall network including all links. (b) Sub-network containing small-weight links with the weights less than the 20\% percentile. (c) Sub-network containing medium-weight links with the weights between the 40\% and 60\% percentiles. (d) Sub-network containing large-weight links with the weights greater than the 80\% percentile.}
    \label{Fig:FertilizerNet:i}
\end{figure*}

The mutual information $I(t, t+1)$ can be suitably normalized by dividing it by the geometric mean of the entropies $H(t)$ and $H(t+1)$ \cite{Strehl-2002,Yao-2003}:
\begin{equation}
  i(t, t+1)=I(t, t+1)/\sqrt{H(t)H(t+1)},
\end{equation}
where $H(t)$ is the entropy of variable $x_t$:
\begin{equation}
\begin{aligned}
  H(t)= &-p_{\mathscr{G}(t)}(0) \log p_{\mathscr{G}(t)}(0)  \\ 
        &-p_{\mathscr{G}(t)}(1) \log p_{\mathscr{G}(t)} (1),
  \end{aligned}
\end{equation}
and $H(t+1)$ is the entropy of variable $x_{t+1}$:
\begin{equation}
  \begin{aligned}
  H(t+1)= &-p_{\mathscr{G}(t+1)}(0) \log p_{\mathscr{G}(t+1)}(0) \\         &-p_{\mathscr{G}(t+1)}(1) \log p_{\mathscr{G}(t+1)} (1).
  \end{aligned}
\end{equation} 
We note that the normalized mutual information $i(t, t+1)$ is equal to 1 if the two successive networks $\mathscr{G}(t)$ and $\mathscr{G}(t+1)$ are identical.


We plot the evolution of the normalized mutual information $i(t,t+1)$ of overall network containing all links in Fig.~\ref{Fig:FertilizerNet:i}(a). As time grows, the normalized mutual information $i(t,t+1)$ maintains an growth trend. The normalized mutual information $i(t,t+1)$ between $\mathscr{G}(1991)$ and $\mathscr{G}(1992)$ has an obvious decrease which was caused by the disintegration of the Soviet Union. The 2007-2009 food crisis and the 2008-2010 economic crisis also cause a significant decrease in the values of $i(2007,2008)$ and $i(2008,2009)$, but correspondingly, the network structures survived the crisis show great stability in the subsequent years.

When we compare the results of the small-, medium- and large-weight sub-networks illustrated in Fig.~\ref{Fig:FertilizerNet:i}(b-d), we find that the normalized mutual information is high for large-weight sub-networks. In addition, there is a sharp drop in the small- and medium-weight sub-networks in late 1990s followed by a rapid recovery. The sharp drop is caused by the Asian financial crisis and the Russian crisis.

\begin{figure*}[htb]
    \centering
    \includegraphics[width=0.49\linewidth]{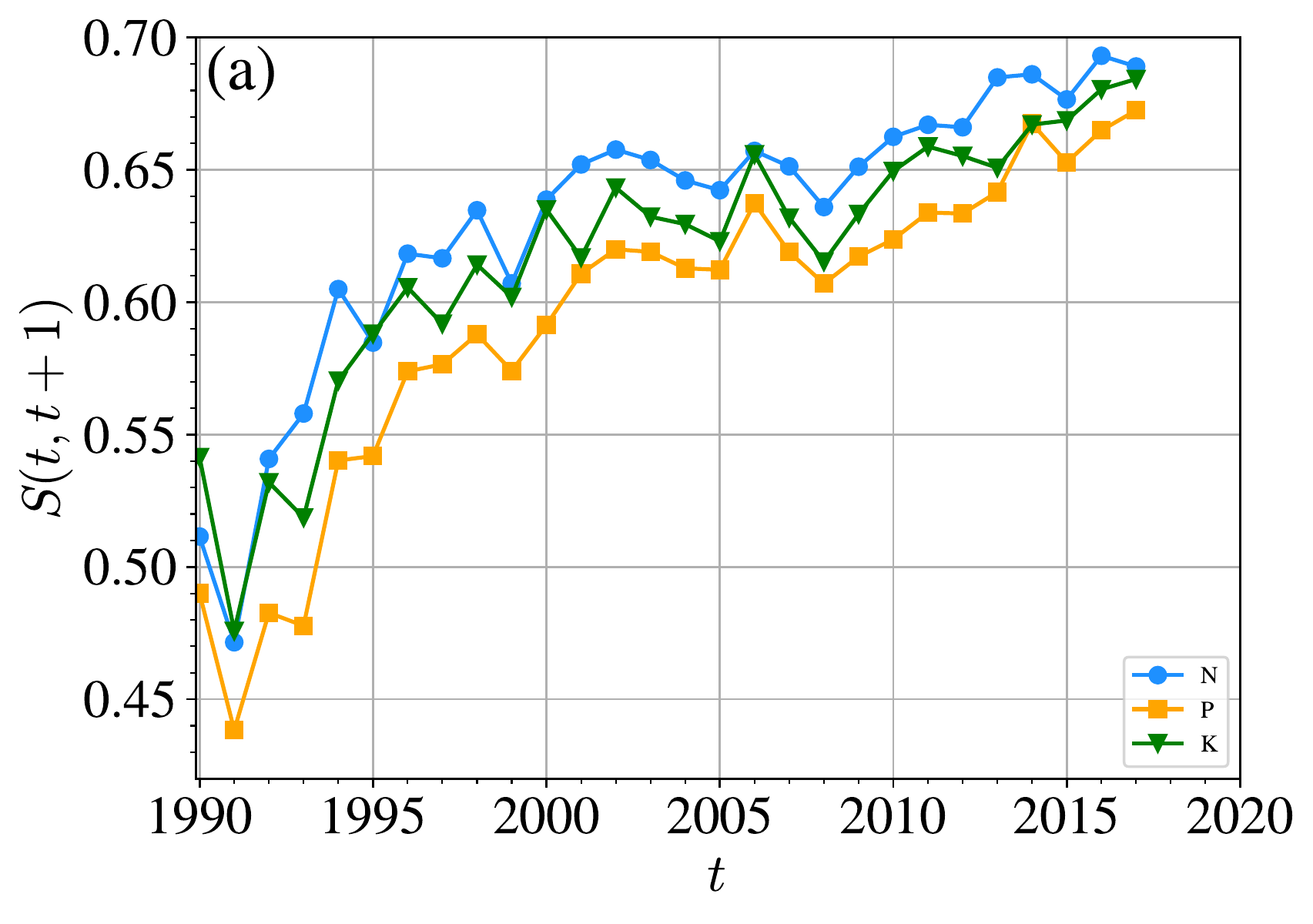}
    \includegraphics[width=0.49\linewidth]{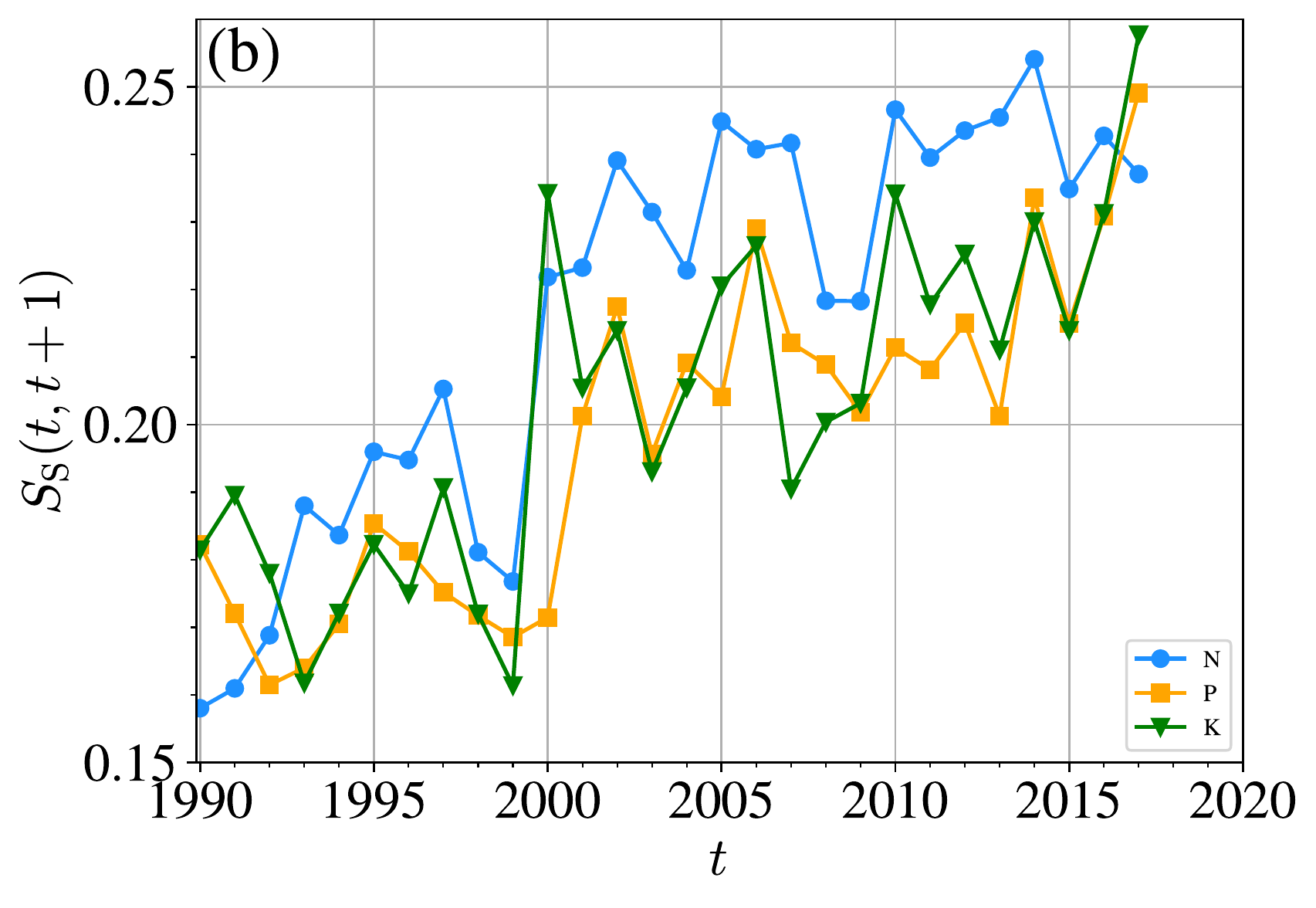}\\
    \includegraphics[width=0.49\linewidth]{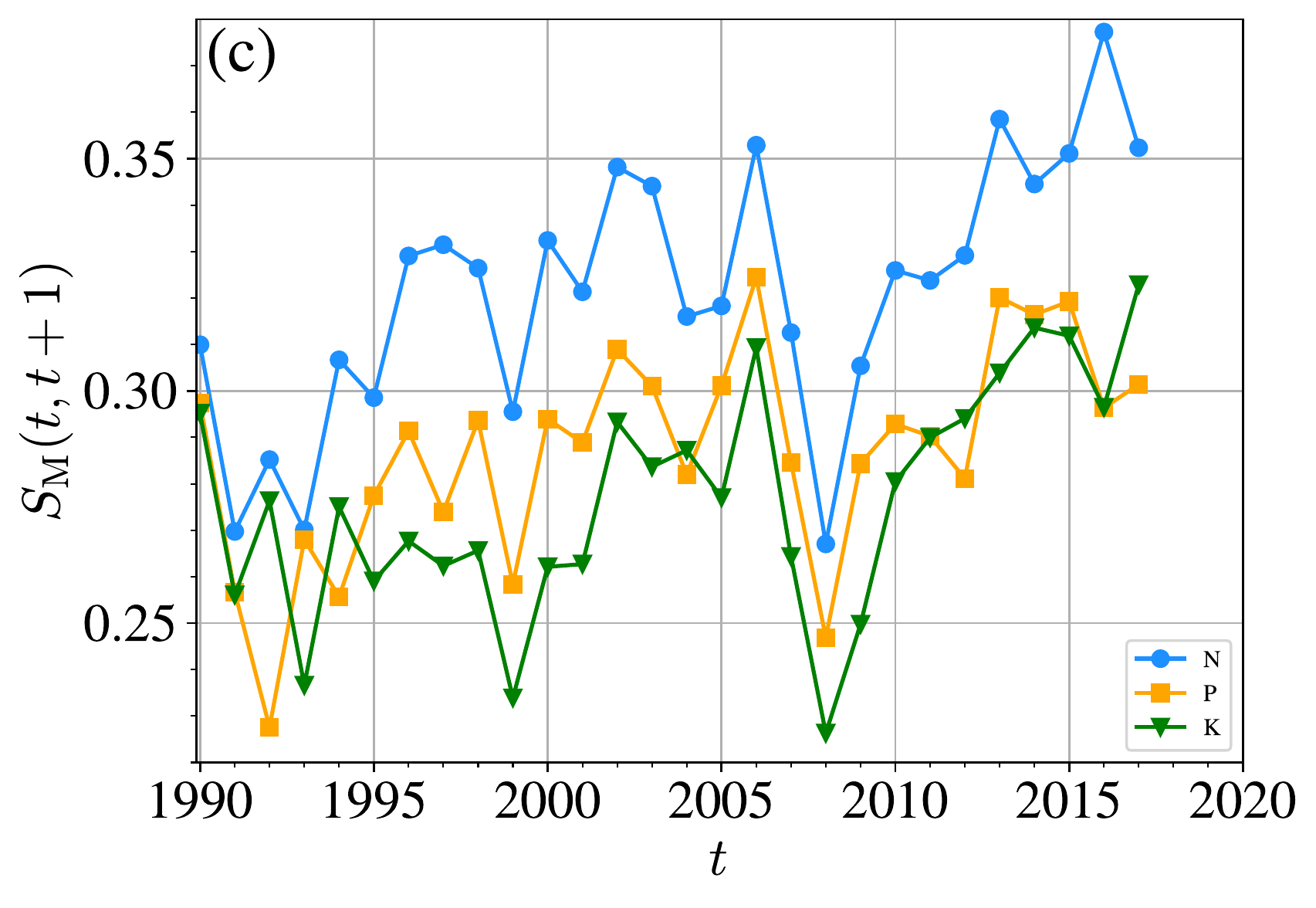}
    \includegraphics[width=0.49\linewidth]{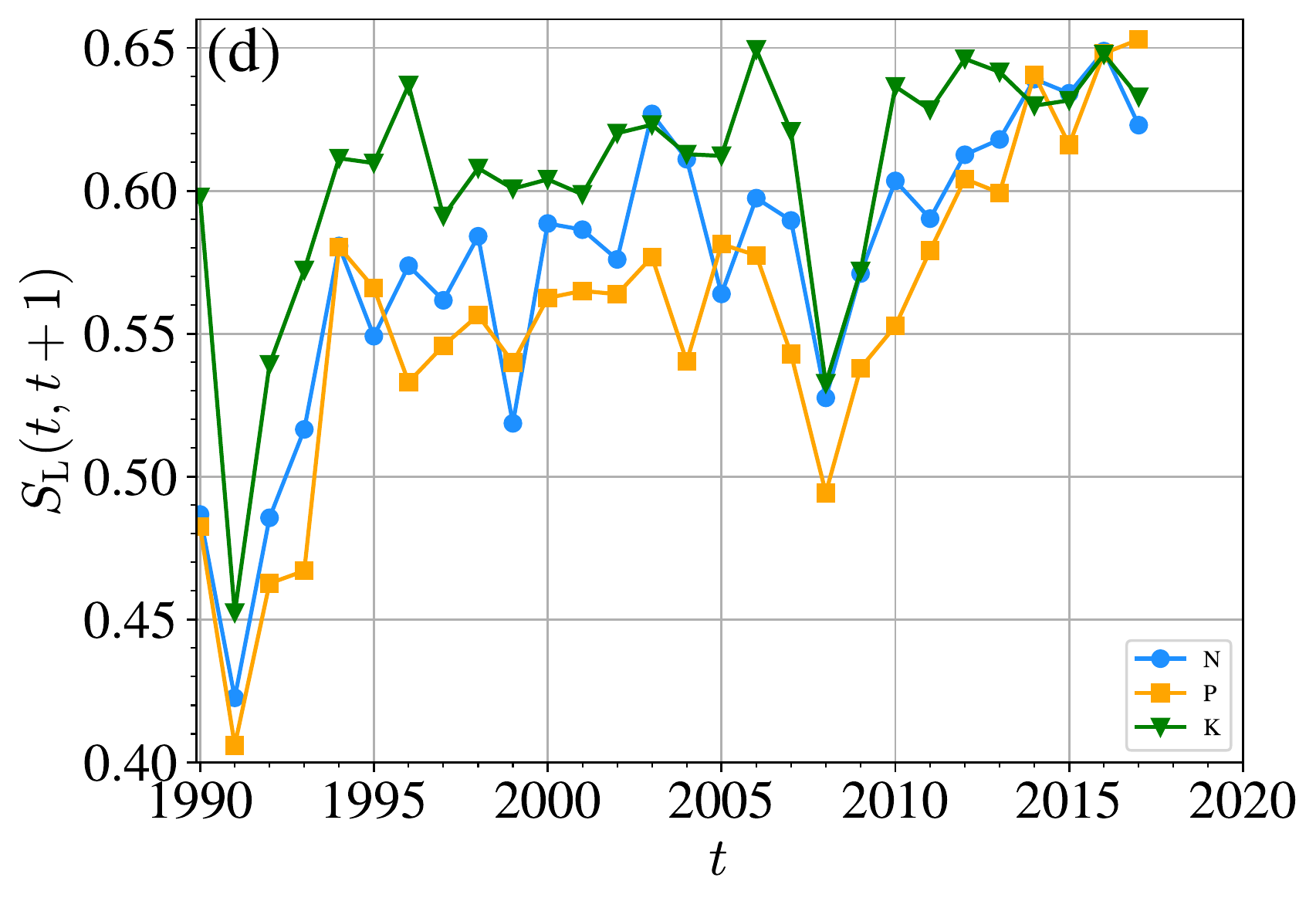}
    \caption{Evolution of the temporal similarity coefficient $S(t,t+1)$ between two successive networks of the international N, P and K trade from 1990 to 2018. (a) Overall network containing all links. (b) Sub-network containing small links with the weights less than the 20\% percentile. (c) Sub-network containing medium links with the weights between the 40\% and 60\% percentiles. (d) Sub-network containing large links with the weights greater than the 80\% percentile.}
    \label{Fig:FertilizerNet:Similarity:t}
\end{figure*}

Comparing the evolution trends in Fig.~\ref{Fig:FertilizerNet:i} and Fig.~\ref{Fig:FertilizerNet:rho}, it is hard to say the two are very similar. For well-known event shocks, both indicators reflect the same impact to convince the indicator's validity. The difference between $r(t,t+1)$ and $i(t,t+1)$ allows us to recognize the stability of the temporal network from the two dimensions of inheritance and correlation. It can be considered as a deeper characterization to reflect
the objective stability of trade networks.

\subsection{Jaccard index}

In order to compare the temporal stability of two successive networks at $t$ and $t+1$ completely, we calculate the similarity coefficient as a dimension. When any two successive networks are more similar, the stability of the temporal network is higher. Here we use the Jaccard index as a direct measure of temporal similarity.

The Jaccard index between two successive networks ${\mathscr{G}}(t)$ and ${\mathscr{G}}(t+1)$ is defined as the ratio of the number of overlapping directed links in the two networks and the number of all directed links appeared in the two networks:
\begin{equation}
    S(t, t+1) 
    = \frac{\sharp\left[{\mathscr{E}}(t)\cap {\mathscr{E}}(t+1)\right]} {\sharp\left[{\mathscr{E}}(t)\cup {\mathscr{E}}(t+1)\right]}.
    \label{Eq:NetSim:Overlap}
\end{equation}
where ${\mathscr{E}}(t)\cup{\mathscr{E}}(t+1)$ is the union set of directed links and ${\mathscr{E}}(t)\cap{\mathscr{E}}(t+1)$ is the intersection of directed links in two successive networks ${\mathscr{G}}(t)$ and ${\mathscr{G}}(t+1)$ defining the set of overlap or coincidence links in both networks. 
The similarity coefficient $S(t, t+1)$ takes a value in $[0,1]$. When the two networks have exactly the same set of directed links, that is, ${\mathscr{E}}(t)={\mathscr{E}}(t+1)$, we have $S(t)=1$. When the two networks have no directed links in common, that is, ${\mathscr{E}}(t)\cap{\mathscr{E}}(t+1)=\emptyset$, we have $S(t, t+1)=0$. The similarity coefficient defined here is not the same as the temporal correlation coefficient \cite{Tang-Scellato-Musolesi-Mascolo-Latora-2010-PhysRevE,Buttner-Salau-Krieter-2016a-SpringerPlus,Buttner-Salau-Krieter-2016b-SpringerPlus}, which is an average of temporal correlation coefficients of nodes in successive networks. 

Fig.~\ref{Fig:FertilizerNet:Similarity:t}(a) shows the evolution of the temporal similarity coefficient $S(t,t+1)$ between two successive networks of overall N, P and K trade from 1990 to 2018. The results are similar to those of the mutual information $i(t,t+1)$ presented in Fig.~\ref{Fig:FertilizerNet:i}. The similarity coefficient increases fast first and then grows slowly, accompanied by several drops. The N trade networks have the highest similarity and the P trade networks have the lowest similarity. 

We calculate the similarity coefficients of sub-networks containing small-weight links with the weights less than the 20\% percentile in Fig.~\ref{Fig:FertilizerNet:Similarity:t}(b), medium-weight links with the weights between the 40\% and 60\% percentiles in Fig.~\ref{Fig:FertilizerNet:Similarity:t}(c) and large-weight links with the weights greater than the 80\% percentile in Fig.~\ref{Fig:FertilizerNet:Similarity:t}(d). The results show that large international trade relationships have higher similarity than small international trade relationships. It means that the network structure spanned by large weights in ${\mathscr{G}}(t)$ is more like the corresponding structure in ${\mathscr{G}}(t+1)$. The N trade networks have the highest similarity in the sub-networks of medium-weight links. In contrast, in the sub-networks of large-weight links, the K trade networks show the highest similarity. The similarity coefficients of medium-weight links and large-weight links experience a huge decline between ${\mathscr{G}}(2008)$ and ${\mathscr{G}}(2009)$, because of the impact of the food and financial crises, while the changes of small trade are not that big.

To sum up, in the results of the three measurement methods, the results of mutual information $i(t,t+1)$ and Jaccard index $S(t,t+1)$ are very similar, indicating that correlation and similarity are strongly related. Although the corresponding physical meanings are not the same, the numerical differences of the two indicators are small. The result of structural inheritance $r(t,t+1)$ is quite different from the two, describing the evolution of the system in a multi-dimensional manner. All three indicators accurately reflect the impact of large events, indicating that they can effectively reflect the objective facts of the trading system.

\section{Conclusion}
\label{S:Conclusion}

In this paper, we use three different and representative methods, structural inheritance, mutual information and Jaccard index, to quantify the temporal stability of international trade networks, taking three fertilizers as the examples. The three measurements have their own physical meanings to evaluate the temporal stability of the network in multiple dimensions. Structural inheritance $r$ refers to the tendency of the structure in ${\mathscr{G}}(t)$ to retain in ${\mathscr{G}}(t+1)$. 
For the correlation between the two networks, we use mutual information $i$ to characterize, which is the uncertainty of the structure of ${\mathscr{G}}(t+1)$ reduced by the structure of ${\mathscr{G}}(t)$. 
The Jaccard index $S$ is the similarity between ${\mathscr{G}}(t)$ and ${\mathscr{G}}(t+1)$, obtained by measuring the ratio of overlapping directed links. 
We use three indicators to measure the structural differences between ${\mathscr{G}}(t)$ and ${\mathscr{G}}(t+1)$ to quantify the temporal stability of the international N, P and K trade systems.

The evolution of the N, P and K trade networks from 1990 to 2018 is estimated by three measurements, from which we find that the stability of the trade system has an upward trend over time. This means that the trade system is becoming more and more mature and stable, and the trade network's ability to maintain its own structure is improving. The evolution process of the three indicators clearly reflects the different influences of historical extreme events on them. The overall trend of each indicator of the three nutrients is similar. 

In the evolution of the three indicators of the sub-network segmented by link weight, we find that the trade system presents a core-periphery-like structure. The core area has strong stability and can maintain its own stability well against time changes. However, it will inevitably suffer huge impacts in the face of large shocks. On the contrast, the stability of the periphery area is weak, but the impact of big events it receives is low because of its nonstop break-born status. Comparing the results of N, P and K trade, most valued N trade has the most mature transit system with the highest stability in the non-core area. P trade is least stable in the early stage, but has caught up in recent years. The K trade network has the most powerful stable center but less stable in other regions due to its high dependence on ore resources.

As the temporal network can describe the evolution of actual complex system in detail, the temporal stability characterizes some interesting characteristics of the system. Taking the temporal networks of N, P and K trade as examples, we use three metrics to comprehensively measure the temporal stability and reveal inner structure. The methods we use can be broadly applied to the study of other temporal complex networks.

\section{Acknowledgements}

This work was partly supported by the National Natural Science Foundation of China (72171083), the Shanghai Outstanding Academic Leaders Plan, and the Fundamental Research Funds for the Central Universities.

\bibliography{Bib1,Bib2,BibITN,BibRobustNet,BibRCE}


\end{document}